\documentclass[aoas,MSNbibl,nameyear,rotating,dvips]{arximspdf}
\usepackage{dcolumn}
\usepackage{graphicx}

\doi{10.1214/13-AOAS702} %kopijuoti is PTS
\volume{8}
\issue{2}
\pubyear{2014}
\firstpage{852}
\lastpage{885}

\makeatletter
\newcolumntype{d}[1]{D{.}{.}{#1}}
\def\uno\mathbf{1}

\def\SSigma\bolds{\Sigma}

\def\btheta{\bolds{\theta}}
\def\RR{\mathbb{R}}

\def\X{\mathbf{X}}
\def\x{\mathbf{x}}
\def\y{\mathbf{y}}

\def\V\mathbf{V}
\def\cero\mathbf{0}

\newproclaim{remark}{Remark}
\newtheorem{Lemma}{Lemma}
\makeatother

\begin{document}
\begin{frontmatter}

\title{Small area estimation of general parameters with~application to
poverty indicators: A~hierarchical Bayes approach}
\runtitle{Small area estimation of poverty indicators}

\begin{aug}
\author[A]{\fnms{Isabel} \snm{Molina}\corref{}\ead[label=e1]{isabel.molina@uc3m.es}\thanksref{t1,m1}},
\author[B]{\fnms{Balgobin} \snm{Nandram}\ead[label=e2]{balnan@wpi.edu}\thanksref{m2}}
\and
\author[C]{\fnms{J.~N.~K.} \snm{Rao}\ead[label=e2]{balnan@wpi.edu}\thanksref{t2,m3}}
\runauthor{I. Molina, B. Nandram and J.~N.~K. Rao}
\affiliation{Universidad Carlos III de Madrid\thanksmark{m1}, Worcester
Polytechnic Institute\thanksmark{m2} and~Carleton University\thanksmark{m3}}
\address[A]{I. Molina\\
Department of Statistics\\
Universidad Carlos III de Madrid\\
C/Madrid 126, 28903 Getafe, Madrid\\
Spain\\
\printead{e1}} %adresu isvedimo komanda gale!
\address[B]{B. Nandram\\
Department of Mathematical Sciences\\
Worcester Polytechnic Institute\\
100 Institute Road\\
Worcester, Massachusetts 01609\\
USA\\
\printead{e2}}
\address[C]{J.~N.~K. Rao\\
School of Mathematics and Statistics\\
Carleton University\\
Ottawa K1S 5B6\\
Canada\\
\printead{e2}}
\end{aug}
\thankstext{t1}{Supported in part by the Spanish Grants MTM2009-09473
and SEJ2007-64500 from the Spanish Ministerio de Educaci\'on y Ciencia.}
\thankstext{t2}{Supported in part by the Natural Sciences and
Engineering Research Council of Canada.}

% HISTORY:
\received{\smonth{2} \syear{2013}}
\revised{\smonth{11} \syear{2013}}

% ABSTRACT
%
\begin{abstract}
Poverty maps are used to aid important political decisions such as
allocation of development funds by governments and international
organizations. Those decisions should be based on the most
accurate poverty figures. However, often reliable poverty figures
are not available at fine geographical levels or for particular
risk population subgroups due to the sample size limitation of
current national surveys. These surveys cannot cover adequately
all the desired areas or population subgroups and, therefore, models
relating the different areas are needed to ``borrow strength'' from
area to area. In particular, the Spanish Survey on Income and
Living Conditions (SILC) produces national poverty estimates but
cannot provide poverty estimates by Spanish provinces due to the
poor precision of direct estimates, which use only the province
specific data. It also raises the ethical question of whether
poverty is more severe for women than for men in a given province.
We develop a hierarchical Bayes (HB) approach for poverty mapping
in Spanish provinces by gender that overcomes the small province
sample size problem of the SILC. The proposed approach has a wide
scope of application because it can be used to estimate general
nonlinear parameters. We use a Bayesian version of the nested
error regression model in which Markov chain Monte Carlo
procedures and the convergence monitoring therein are avoided. A
simulation study reveals good frequentist properties of the HB
approach. The resulting poverty maps indicate that poverty, both
in frequency and intensity, is localized mostly in the southern
and western provinces and it is more acute for women than for men
in most of the provinces.
\end{abstract}

% KEYWORDS

\end{frontmatter}

%s1 #&#
\section{Introduction}\label{intro}
Before the recent world economic crisis, the goal of a 23\% maximum
global poverty rate established in the United Nations Millennium
Development Goals for the year 2015 seemed to be easy to achieve
and there was also a clear indication of progress in all the other
goals. However, after the crisis and also due to the late
environmental disasters such as the drought in East Africa, the
situation regarding poverty is getting worse. A~reliable and
detailed statistical measurement is certainly essential in the
assessment of the well being of different regions, which will lead
to the design of effective developmental policies.

Often, national surveys are not designed to give reliable
statistics at the local level. This is the case of the Spanish
Survey on Income and Living Conditions (SILC), which is planned
to produce estimates for poverty incidence at the Spanish
Autonomous Communities (large regions), but it cannot provide
estimates for Spanish provinces due to the small SILC sample sizes
for some of the provinces. The population subdivisions, not
necessarily geographical, which constitute the estimation domains,
will be called in general ``areas.'' When estimating some
aggregate characteristic of an area, a ``direct'' estimator is the
one that uses solely the data from that area. These estimators are
often design unbiased, at least approximately. However, they have
overly large sampling errors for areas with small sample sizes.
The areas with inadequate sample sizes are labeled as ``small
areas.'' This problem has given rise to the development of the
scientific field called small area estimation, which studies
``indirect'' estimation methods that ``borrow strength'' from
related areas. Some of these methods are based on explicit models
that link all areas through common parameters and making use of
auxiliary information. Such model-based techniques are appealing
because they provide estimators with high efficiency even under
very small area-specific sample sizes. The monograph of \citet{Rao03} contains a comprehensive account of small area estimation
techniques that appeared until the publication date; see \citet
{JiaLah06}, \citet{Dat09} and \citet{Pfe13} for reviews
of the more recent work.

Small area models may be classified into two broad types:
(i)~Area-level models that relate small area direct estimates to
area-specific covariates, and (ii) Unit-level models that relate the
unit values of a study variable to associated unit-specific
covariates and possibly also area-specific covariates. So far,
most of the model-based small area methods have focused on the
estimation of totals and means, and nonlinear parameters have not
received much attention. However, many poverty and inequality
indicators are rather complex nonlinear functions of the income
or other welfare measures of individuals; see, for example, \citet
{NerBalBet05}. The main purpose of this paper is to
develop a suitable method, based on the hierarchical Bayes
approach, to handle general nonlinear parameters. We, however,
focus on poverty indicators as particular cases due to the
important socio-economic impact of this application.

Hierarchical Bayesian models have been extensively used in small
area estimation; see \citet{Rao03}, Chapter~10 and \citet{Dat09}.
A~hierarchical Bayesian model can accommodate very complex models
for the data based on very simple models as building blocks. For
example, within the Bayesian paradigm, by making parameters
stochastic, one can introduce an intracluster correlation and
different sources of variability can be also incorporated. In
small area estimation, the hierarchical Bayesian model provides
the much needed ``borrowing of strength'' in a simple manner; see, for example,
Nandram and Choi (\citeyear{autokey22}, \citeyear{NanCho10}), where
hierarchical Bayesian
models are used to study body mass index on the continuous scale.
\citet{YouZho11} use a spatial hierarchical Bayes model in an
application to health survey data. Finally, \citet{Mohetal12}
study small area estimation of adult literacy in the U.S. using
unmatched sampling and linking models, using hierarchical Bayes
methods.

In the context of poverty estimation, area-level models have been
used to estimate the proportion of school age children under
poverty at the county level under the SAIPE program (Small Area
Income and Poverty Estimates) of the U.S. Census Bureau; for more
details, see, for example, \citet{Bel} or the program webpage
\url{http://www.census.gov/did/www/saipe/}. Using the Spanish SILC data,
\citet{MolMor09} used an area-level model relating direct
estimates of poverty proportions and poverty gaps (defined in
Section~\ref{povindic}) to area covariates obtained from a much
larger survey, the Labor Force Survey (LFS). In this application,
the areas are the Spanish provinces. Results based on the
empirical best (or Bayes) approach for this area-level model
indicated only marginal gains in efficiency over direct estimates.

Few approaches have appeared in the literature for efficient
estimation of general nonlinear indicators using unit-level
models. Here we discuss the most popular ones. The first one, due
to \citet{ElbLanLan03}, is the method used by the
World Bank (WB). This method was designed specially to deal with
complex nonlinear poverty indicators. It assumes that the log
incomes of the individuals in the population follow a unit-level
model similar to the nested error linear regression model of
\citet{BatHarFul88}, but including random effects
for sampling clusters instead of area effects. After fitting the
model to survey data, the WB method generates by bootstrap resampling
a number of synthetic censuses making use of census auxiliary data
and the fitted model. From each synthetic census, a poverty
indicator of interest is computed for each small area. The average
of the estimates over simulated censuses is then taken as the
point estimate of the poverty indicator, and the variance of the
estimates is taken as a measure of variability. The WB used the above
simple method to produce poverty maps for many countries, by
securing census auxiliary data and income data from a sample
survey. In European countries, registers that provide unit-level
population data may be obtained through collaboration with
statistical offices. In Scandinavian countries and in Switzerland,
continuously updated census auxiliary unit-level data are
available through statistical offices. We emphasize that unit
level auxiliary data for the population is needed to implement the
WB method, based on a unit-level model, to estimate small area
poverty indicators or other complex parameters; area means of the
auxiliary variables would be sufficient in the case of estimating
area means of a variable of interest.

The second approach for estimation of general small area
parameters, based on the empirical best/Bayes (EB) method, was
recently introduced by \citet{MolRao10}. This method gives
estimators with minimum mean squared error called best
predictors or, more exactly, Monte Carlo approximations to the best
predictors. This is done under the assumption that there exists a
transformation of the incomes of individuals or another welfare
variable used to measure poverty such that the transformed incomes
follow the nested error regression model of \citet{BatHarFul88}.
Mean squared errors of the EB estimators are estimated by a
parametric bootstrap method. The method of \citet{MolRao10}
also requires unit-level auxiliary data for the population.

Both methods approximate an expected value by Monte Carlo, which
requires generation of many full synthetic censuses that might
be of huge size (e.g., over 43 million in the Spanish application
of Section~\ref{FreVal}). In addition, the mean squared error is estimated
using the bootstrap (or double bootstrap) in which the expected values
need to be approximated for each bootstrap (or double bootstrap)
replicate. The full procedure might be very intensive
computationally and even may not be feasible for very complex
poverty indicators such as those requiring sorting population
elements or for very large populations such as Brazil or India.

The hierarchical Bayes method is a good alternative to EB because it
does not require the use of bootstrap for mean squared error
estimation and it provides credible intervals and other useful
summaries from the posterior distributions with practically no
additional effort. We propose a very simple hierarchical Bayes
(HB) approach that is computationally much more efficient than
the alternative EB procedure. Only noninformative priors are
considered to save us from the introduction of subjective
information which might be controversial in official statistics
applications. Moreover, using a particular reparameterization of
the model and noninformative priors, we avoid the use of Markov
chain Monte Carlo (MCMC) methods and therefore also the need for
monitoring convergence of the Markov chains for each generated
sample in the simulation studies, but ensuring propriety of the
posterior under general conditions. In our simulations, this HB
method provides point estimates that are practically the same as
EB estimates and inferences that have frequentist validity. This
frequentist validity gives a strong support to the use of the
proposed HB method in practice.

%s2 #&#
\section{Poverty indicators}\label{povindic}
Certainly, poverty and income inequality are broad and complex
concepts which cannot be easily summarized in one measure or
indicator. In the literature there are many different indicators
intending to summarize poverty or income inequality in one
measure, each of them focusing on the measurement of particular
aspects of poverty. For a summary of poverty and inequality
indicators see, for example, \citet{NerBalBet05}. Basic poverty indicators
are the head count ratio, referred to here as poverty incidence,
which is simply the proportion of individuals with welfare measure
under the poverty line, and the poverty gap, measuring the mean
relative distance to the poverty line of the individuals with
welfare measure under the poverty line. The class of poverty
indicators introduced by \citet{FosGreTho84}
contains the previous two as particular cases. Other measures
include the Sen Index, the Fuzzy monetary and the Fuzzy
supplementary poverty indicators [\citet{Betetal06}]. Practically
all poverty measures are rather complex nonlinear functions of
the income or some other welfare measure of individuals. We will
introduce HB methodology that is suitable for the estimation of
general nonlinear parameters, but we illustrate the procedure by
applying it to particular indicators of interest, namely, the
poverty incidence and the poverty gap as in \citet{MolRao10}.
This will allow comparison with the EB method introduced
in that paper.\looseness=-1

Let us consider a population of size $N$ that is partitioned into
$D$ subpopulations of sizes $N_1,\ldots,N_D$ and called small
areas. Let $E_{di}$ be a suitable quantitative measure of welfare
for individual $i$ in small area $d$, such as income or
expenditure, and let $z$ be the poverty line, that is, the
threshold for $E_{di}$ under which a person is considered as
``under poverty.'' The family of poverty measures of \citet{FosGreTho84} for a small area $d$ may be expressed as
\[
F_{\alpha d} =\frac{1}{N_d}\sum_{i=1}^{N_d}
\biggl(\frac{z-E_{di}}{z} \biggr)^{\alpha
}I(E_{di}<z),\qquad\alpha \geq0,  d=1,\ldots,D,
\]
where $I(E_{di}<z)=1$ if $E_{di}<z$ or the person is under poverty
and\break  \mbox{$I(E_{di}<z)=0$} if $E_{di}\geq z$ or the person is not under
poverty. Taking $\alpha=0$, we obtain the area poverty incidence,
which measures the frequency of the poverty, and $\alpha=1$
leads to the area poverty gap, which quantifies the intensity of
the poverty.

%s3 #&#
\section{Hierarchical Bayes predictors of poverty indicators}\label{HBmethod}
Estimation of the target area characteristics is based on a
random sample drawn from the finite population according to a
specified sampling design. Let $P$ denote the set of indices of
the population units, $s$ be the set of units selected in the
sample, of size $n<N$, and $r=P-s$ be the set of the units, with
size $N-n$, that are not selected. Let $P_d$, $s_d$, $r_d$, $N_d$
and $n_d$ be, respectively, the set of population units, sample,
sample complement, population size and sample size, restricted to
area $d$. We allow zero sample sizes for some of the areas.
Without loss of generality, we assume that those areas are the
last $D-D^*$ areas, that is, $n_d>0$, for $d=1,\ldots,D^*$, where
$D^*\leq D$, and $n_d=0$ for $d=D^*+1,\ldots,D$. Then the overall
sample size is $n=n_1+\cdots+n_{D^*}$. For $d=D^*+1,\ldots,D$ with
$n_d=0$, we have $s_d=\varnothing$ and $r_d=P_d$.

To estimate $F_{\alpha d}$ efficiently for each area $d$, we assume
that there are $p$ auxiliary variables related linearly to some
one-to-one transformation $Y_{di}=T(E_{di})$ of the welfare variables.
More concretely, we assume that
the transformed population values $\{Y_{di};i=1,\ldots,N_d\}$
follow the nested error model
%
%e1 #&#
\begin{equation}
\label{origNEmodel} Y_{di}=\x_{di}'\bolds{
\beta}+u_d+e_{di},\qquad i=1,\ldots,N_d, d=1,\ldots,D,
\end{equation}
introduced by \citet{BatHarFul88}, where $\x_{di}$ is the
$p\times1$
vector of auxiliary variables for unit $i$ within area $d$, $\bolds
{\beta}$ is the
$p\times1$ (constant) vector of regression coefficients associated
with $\x_{di}$, $u_d$ is a random effect of area
$d$, which models the unexplained between area variation, and
$e_{di}$ is the individual model error. Area effects $u_d$ and errors
$e_{di}$ given all parameters are independent and
satisfy, respectively, $u_d|\sigma_u^2\stackrel{\mathrm{i.i.d.}}\sim
N(0,\sigma
_u^2)$ and
$e_{di}|\sigma^2\stackrel{\mathrm{i.i.d.}}\sim N(0,\sigma
^2w_{di}^{-1})$, where
$w_{di}>0$ is a known heteroscedasticity
weight. In practice, these weights can be obtained from a preliminary
modeling of
error variances using variables different from those considered in the
mean model as done in the WB method.
We assume that the values of the auxiliary variables are known for all
population units.

{MCMC is a popular tool used to implement the HB method and
software such as WinBUGS is readily available. However, running
the Gibbs sampler requires monitoring the convergence by making a
long run, thinning and performing convergence tests. In simulation
studies, this monitoring process must be done for each simulated
data set. Failing to do this carefully might lead to gross
approximation of the desired quantities. Instead, making random
draws directly from the posterior whenever possible avoids the
need for monitoring the convergence and can therefore save a
considerable amount of time.} Here we consider a particular
reparameterization of the model which, together with
noninformative priors, provides a generally proper posterior,
and at the same time a way to skip MCMC procedures by randomly
drawing from the posterior distribution using the chain rule of
probability. The new reparameterization is based on expressing
the model in terms of the intra-class correlation
$\rho=\sigma_u^2/(\sigma_u^2+\sigma^2)$ as
%
%e2 #&#
%e3 #&#
\begin{eqnarray}
Y_{di}|u_d,\bolds{\beta},\sigma^2
&\stackrel{\mathrm{ind}}\sim& N\bigl(\x_{di}'\bolds{
\beta}+u_d,\sigma^2w_{di}^{-1}\bigr),
\label{condit}
\\
 u_d|\rho,\sigma^2 &\stackrel{\mathrm{ind}}\sim& N \biggl(0,
\frac{\rho}{1-\rho} \sigma^2 \biggr), \qquad i=1,\ldots,
N_d, d=1,\ldots,D, \label{ud}
\end{eqnarray}
see, for example, \citet{TotNan10} for a similar formulation of
the nested error model (\ref{origNEmodel}).

We assume that the population model given by (\ref{condit}) and (\ref
{ud}) holds for the sample units
$s_d$ and for the out-of-sample units $r_d$; that is, the sampling
design is noninformative and therefore sample selection bias is
absent. We may also point out that the WB method implicitly
assumes that the model fitted for the sample data also holds for
the population in order to generate synthetic censuses of the
variable of interest. \citet{PfeSve07} considered
the estimation of small area means under informative sampling in
the context of two-stage sampling. This method requires the
modeling of sampling weights in terms of the variable of interest
and auxiliary variables. It is not clear how this method may be
extended to handle complex parameters such as poverty indicators
and to other sampling designs. In the application with data from the
Spanish SILC
described in Section~\ref{applic}, we provide
graphical diagnostics to check for informative sampling.

Note that the untransformed welfare variables $E_{di}$ can be obtained
from the model responses
as $E_{di}=T^{-1}(Y_{di})$, where $T^{-1}(\cdot)$ denotes the inverse
transformation of $T(\cdot)$.
Then, the FGT poverty indicator $F_{\alpha d}$ is a nonlinear function of
the vector $\y_d=(Y_{d1},\ldots,Y_{dN_d})'$ of response variables for
area $d$. Thus, more generally, our aim is to estimate
through the HB approach a general area parameter $\delta_d=h(\y_d)$,
where $h(\cdot)$ is a measurable function.

In the case of estimating particular area parameters $\delta_d$ that
have social relevance such as poverty indicators or when the results are
going to aid political decisions, the introduction of subjective
informative priors might not be acceptable. For this reason,
here we consider only noninformative priors for the unknown model
parameters $(\bolds{\beta}',\sigma^2,\rho)$.
Consider the following simpler situation, without covariates and with
only one observation:
\[
Y | \mu\sim N\bigl(\mu, \sigma^2\bigr),\qquad \mu\sim N\bigl(\theta,
\delta^2\bigr).
\]
Now, let us define the intraclass correlation $\rho=
\delta^2/(\delta^2 + \sigma^2)\in(0,1)$. By Bayes' theorem, the
posterior density of $\mu$ is $\mu| Y \sim N\{\rho Y +
(1-\rho)\theta, \rho\sigma^2\}$, which leads to the shrinkage or
reference prior for $\rho$ given by $\rho\sim U(0,1)$; see
\citet{NatKas00}. Shrinkage priors lead to good
frequentist properties of HB inferences. In our model, to ensure
propriety of the posterior of
$\rho=\sigma_u^2/(\sigma_u^2+\sigma^2)$, we consider a uniform
prior for $\rho$ in any closed interval of $(0,1)$, that is, in
$[\varepsilon,1-\varepsilon]$, $\varepsilon>0$. In practice, taking
$\varepsilon=0.0001$ should suffice; see, for example, Figure~\ref{PostRhoMales}. Next, consider the simpler model
\[
Y | \sigma^2 \sim N\bigl(\mu, \sigma^2\bigr).
\]
Under this model, Jeffreys' reference prior for $\sigma^2$ is
$\pi(\sigma^2) \propto1/\sigma^2$, $\sigma^2 > 0$.
{Jeffreys' prior is said to be objective because it is the square
root of Fisher's information. It has three important
properties. First, it is invariant to one-to-one \mbox{transformations}
of the parameter, which makes it convenient for scale parameters.
Second, it is constructed using only the likelihood function and
no other subjective judgement is needed. Third, it typically does
not involve other hyperparameters requiring the specification of
further priors. For these reasons, Jeffreys' prior is widely
accepted in the literature.} Thus, for the unknown parameters
$(\bolds{\beta}',\sigma^2,\rho)$ in~model (\ref{condit})--(\ref{ud}), we
consider the noninformative prior
%
%e4 #&#
\begin{equation}
\label{prior} \pi\bigl(\bolds{\beta},\sigma^2,\rho\bigr) \propto
\frac{1}{\sigma^2},\qquad{\varepsilon\leq\rho\leq1-\varepsilon,
\sigma^2>0,  \bolds{\beta} \in\RR^p}.
\end{equation}

Let $\mathbf{u}=(u_1,\ldots,u_D)'$ be the vector of random area effects and
$\y=(\y_1',\ldots,\break \y_D')'$ the vector containing all the population
response variables.
Sorting by sample and out-of-sample units, this vector can be
expressed as
$\y=(\y_s',\y_r')'$, where $\y_s$ contains the elements of $\y$ corresponding
to sample units and $\y_r$ to out-of-sample units.
For convenience, we will use the notation $\btheta=(\mathbf{u}',\bolds
{\beta}',\sigma^2,\rho)$.
The above choice of priors allows us to avoid MCMC by using the chain
rule of probability to represent the joint
posterior density of $\btheta$ as follows:
%
%e5 #&#
\begin{eqnarray}\label{post}
&& \pi\bigl(\mathbf{u},\bolds{\beta},\sigma^2,\rho|\y_s\bigr)
\nonumber\\[-8pt]\\[-8pt]
&&\qquad =\pi_1\bigl(\mathbf{u}|\bolds{\beta},\sigma^2,\rho,\y_s\bigr)
\pi_2\bigl(\bolds{\beta}|\sigma^2,\rho,\y_s
\bigr) \pi_3\bigl(\sigma^2|\rho,\y_s\bigr)
\pi_4(\rho|\y_s).\nonumber
\end{eqnarray}
Here, $\pi_1(\mathbf{u}|\bolds{\beta},\sigma^2,\rho,\y_s)$, $\pi
_2(\bolds{\beta}|\sigma^2,\rho,\y_s)$ and
$\pi_3(\sigma^2|\rho,\y_s)$ in (\ref{post}) have simple closed forms,
but $\pi_4(\rho|\y_s)$ is not simple; see Appendix~\ref{posterior}. However, random
values from
$\pi_4(\rho|\y_s)$ can be drawn using a grid method or an
accept--reject algorithm; see Section~\ref{FreVal}. A similar drawing
procedure using
the chain rule was mentioned in \citet{Ber85}, Section~4.6.
\citet{DatGho91} also used this analytical approach for HB
estimation of small area means
under linear mixed models, but employing gamma priors on the
reciprocals of variance components.
Lemma~\ref{le1app2} in Appendix~\ref{propriety} states that the posterior density
in (\ref{post}) is proper provided that the matrix $X=\mathrm{col}_{1\leq d\leq D}\mathrm{col}_{i\in s_d} (\x_{di}')$ has full column
rank and $\varepsilon\leq\rho\leq1-\varepsilon$, $\varepsilon>0$.

Now since the model (\ref{condit}) holds for all the population units,
given the vector of parameters
$\btheta$ which includes area effects, out-of-sample responses $\{
Y_{di},i\in r_d\}$ are independent of sample responses $\y_s$ with
%
%e6 #&#
\begin{equation}
\label{predfdi} Y_{di}|\btheta\stackrel{\mathrm{ind}}\sim N\bigl(
\x_{di}'\bolds{\beta}+u_d,
\sigma^2w_{di}^{-1}\bigr),\qquad i\in
r_d,  d=1,\ldots,D.
\end{equation}
Consider the sample and out-of-sample decomposition of the area vector
$\y_d=(\y_{ds}',\y_{dr}')'$.
The posterior predictive density of $\y_{dr}$ is given by
\[
f(\y_{dr}|\y_s)=\int\prod_{i\in r_d}
f(Y_{di}|\btheta)\pi(\btheta|\y_s)\,d\btheta.
\]
The HB estimator of the target parameter $\delta_d=h(\y_d)$ is then
given by the posterior mean
\[
\hat\delta_d^{\mathrm{HB}} =E(\delta_d|
\y_s)=\int h(\y_{ds},\y_{dr})f(\y_{dr}|
\y_s)\,d \y_{dr},
\]
which can be approximated by Monte Carlo. This approximation is
obtained by first generating samples from the posterior
$\pi(\btheta|\y_s)$. For this, we first draw $\rho$ from
$\pi_4(\rho|\y_s)$, then $\sigma^2$ from $\pi_3(\sigma^2|\rho,\y_s)$,
then $\bolds{\beta}$ from
$\pi_2(\bolds{\beta}|\sigma^2,\rho,\y_s)$ and finally $\mathbf{u}$
from $\pi_1(\mathbf{u}|\bolds{\beta},\sigma^2,\rho,\y_s)$.
We can repeat this procedure a large number, $H$, of times to get a
random sample
$\btheta^{(h)}$, $h=1,\ldots,H$ from $\pi(\btheta|\y_s)$. Then, for
each generated
$\btheta^{(h)}$, $h=1,\ldots, H$, from $\pi(\btheta|\y_s)$, we draw
out-of-sample values
$Y_{di}^{(h)}$, $i\in r_d$, $d=1,\ldots,D$, from the distribution in
(\ref{predfdi}).
Thus, for each sampled area $d=1,\ldots,D^*$, we have generated an
out-of-sample vector
$\y_{dr}^{(h)}=\{Y_{di}^{(h)}, i\in r_d\}$ and we have also
the sample data $\y_{ds}$ available. Thus, we construct the full
population vector
$\y_d^{(h)}=(\y_{ds}',(\y_{dr}^{(h)})')'$.

For each nonsampled area $d=D^*+1,\ldots,D$, the whole vector $\y
_d^{(h)}=\y_{dr}^{(h)}$
is generated from (\ref{predfdi}) since in that case $r_d=P_d$.
Using $\y_d^{(h)}$, we compute the area parameter $\delta_d^{(h)}=h(\y
_d^{(h)})$, $d=1,\ldots,D$.
In the particular case of estimating the FGT poverty measure $\delta
_d=F_{\alpha d}$,
using $\y_d^{(h)}$, we calculate
%
%e7 #&#
\begin{equation}
\label{MCF}  F_{\alpha d}^{(h)} =\frac{1}{N_d} \biggl[\sum
_{i\in s_d} \biggl(\frac{z-E_{di}}{z} \biggr)^{\alpha}I(E_{di}<z)
+ \sum_{i\in
r_d} \biggl(\frac{z-E_{di}^{(h)}}{z}
\biggr)^{\alpha
}I\bigl(E_{di}^{(h)}<z\bigr) \biggr],\hspace*{-35pt}
\end{equation}
where $E_{di}=T^{-1}(Y_{di})$, $i\in s_d$ and
$E_{di}^{(h)}=T^{-1}(Y_{di}^{(h)})$, $i\in r_d$, $d=1,\ldots,D$.
Thus, in this way we have a random sample $\delta_d^{(h)}$, $h=1,\ldots,H$, from the posterior density
of the target parameter $\delta_d$. Finally, the HB estimator $\hat
\delta_d^{\mathrm{HB}}$, under squared loss, is the\vadjust{\goodbreak} posterior
mean obtained by averaging $\delta_d^{(h)}$ over $h=1,\ldots,H$.
As an uncertainty measure, we consider the posterior variance obtained
as the variance of the $\delta_d^{(h)}$ values.
Thus,
%
%e8 #&#
\begin{equation}
\label{MCest} \hat\delta_d^{\mathrm{HB}}=E(\delta_d|
\y_s)\approx\frac{1}{H}\sum_{h=1}^H
\delta_d^{(h)}, \qquad V(\delta_d|
\y_s)\approx\frac{1}{H}\sum_{h=1}^H
\bigl(\delta_d^{(h)}-\hat\delta_d^{\mathrm{HB}}
\bigr)^2.
\end{equation}
Other useful posterior summaries such as credible intervals can be
computed in a~straightforward
manner.
%
%re1 #&#
\begin{remark}
When the target area parameter is computationally complex, such as
indicators based on
pairwise comparisons or sorting area elements, or when the population
is too large,
a faster HB approach can be implemented analogously to the fast
EB\vadjust{\goodbreak}
approach introduced in \citet{FerMol12}.
For this, from each Monte Carlo population vector $\y_d^{(h)}$ we draw
a sample $s_d^{(h)}$ using the original sampling design and, with this
sample, we obtain
a design-based estimator $\hat\delta_d^{(h)}$ of~$\delta_d^{(h)}$.
This value would replace $\delta_d^{(h)}$ in (\ref{MCest}), that is,
the estimator would be given by $\hat\delta_d^{\mathrm{FHB}}=H^{-1}\sum_{h=1}^H
\hat\delta_d^{(h)}$. The posterior variance can
be approximated similarly by $H^{-1}\sum_{h=1}^H (\hat\delta
_d^{(h)}-\hat\delta_d^{\mathrm{FHB}} )^2$.
\end{remark}

%s4 #&#
\section{Model validation}
In practice, results based on a model should be validated by analyzing
how good the assumed model fits our data.
Under the HB setup, several validation measures have been
proposed in the literature. Here we consider the cross-validation approach
advocated by \citet{GelDeyCha92}, based on looking at
the predictive distribution of each observation when that observation
has been
deleted from the sample. As validation statistics, we consider
the standardized cross-validation residuals used in a similar model to
ours by
\citet{NanSedPic00} and the conditional predictive ordinates
defined by \citet{B80} and studied under normal distributions by
\citet{Pet90}.

Standardized cross-validation residuals are defined as
%
%e9 #&#
\begin{equation}
\label{delres} r_{di}=\frac{Y_{di}-E (Y_{di}|\y_{s(di)} )}{\sqrt
{V(Y_{di}|\y
_{s(di)})}}, \qquad i\in s_d, d=1,\ldots,D,
\end{equation}
where $\y_{s(di)}$ is the data vector excluding observation
$Y_{di}$. Recently, \citet{Wanetal12} used these residuals for a
similar assessment on an agricultural
application. Interpretation of diagnostic plots obtained using these
residuals needs to be
cautious because by construction they are correlated.
However, in the application of Section~\ref{applic}, diagnostic
plots using these residuals look practically the same as those
obtained from the usual
frequentist residuals delivered by a maximum likelihood fit of the
original nested error regression model
(\ref{origNEmodel}). In the remainder of this section we
explain how to obtain Monte Carlo approximations of the expected value
and variance in (\ref{delres}).

Following \citet{GelDeyCha92}, if we generate $H$ independent values
$\btheta^{(h)}= ((\mathbf{u}^{(h)})',(\bolds{\beta}^{(h)})',\sigma
^{2(h)},\rho^{(h)})'$, $h=1,\ldots,H$, from the posterior
density given all the data, $\pi(\btheta|\y_s)$, the posterior
expectation in (\ref{delres}) can be approximated by a weighted average as
\begin{eqnarray*}
E (Y_{di}|\y_{s(di)} ) &=& \int E (Y_{di}|\y
_{s(di)},\btheta)\pi(\btheta|\y_{s(di)}) \,d\btheta
\\
&=& \int\bigl\{\x_{di}'\bolds{\beta}+u_d
\bigr\}\pi(\btheta|\y_{s(di)}) \,d\btheta
\\
&\approx& \sum_{h=1}^H \bigl\{
\x_{di}'\bolds{\beta}^{(h)}+u_d^{(h)}
\bigr\} v_{di}^{(h)},\qquad i\in s_d, d=1,\ldots,D.
\end{eqnarray*}
Here, the weights $v_{di}^{(h)}$ are given by
%
%e10 #&#
\begin{equation}
\label{vdi} v_{di}^{(h)}= \Biggl[f\bigl(Y_{di}|
\btheta^{(h)}\bigr)\sum_{k=1}^H
\bigl\{ f\bigl(Y_{di}|\btheta^{(k)}\bigr) \bigr
\}^{-1} \Biggr]^{-1},
\end{equation}
where $f(Y_{di}|\btheta)$ is the normal density indicated in (\ref{condit}).
This has been obtained from the fact that, given $\btheta$, all
observations are independent and
distributed as indicated in (\ref{predfdi}), using Bayes' theorem and
taking into account that
$f(\y_s|\btheta)=f(\y_{s(di)}|\btheta)f(Y_{di}|\btheta)$; for more
details see
Appendix~\ref{Appendix3}. To obtain the posterior variance $V(Y_{di}|\y
_{s(di)})=E(Y_{di}^2|\y_{s(di)})-E^2(Y_{di}|\y_{s(di)})$, the
expectation $E(Y_{di}^2|\y_{s(di)})$ can be
approximated similarly, by
\begin{eqnarray*}
E \bigl(Y_{di}^2|\y_{s(di)} \bigr) &=& \int E
\bigl(Y_{di}^2|\y_{s(di)},\btheta\bigr)\pi(\btheta|
\y_{s(di)}) \,d\btheta
\\
&=& \int\bigl\{\sigma^2w_{di}^{-1}+\bigl(
\x_{di}'\bolds{\beta}+u_d\bigr)^2
\bigr\}\pi(\btheta|\y_{s(di)}) \,d\btheta
\\
&\approx& \sum_{h=1}^H \bigl\{
\sigma^{2(h)}w_{di}^{-1}+\bigl(\x_{di}'
\bolds{\beta}^{(h)}+u_d^{(h)}\bigr)^2
\bigr\} v_{di}^{(h)},\qquad i\in s_d.
\end{eqnarray*}

To further assess the model, we use the conditional predictive
ordinates (CPOs).
For observation $Y_{di}$, the CPO is defined as
the predictive density of $Y_{di}$ given the sample data with that
observation deleted, that is,
\[
\mathrm{CPO}_{di}=f(Y_{di}|\y_{s(di)})=\int
f(Y_{di}|\y_{s(di)},\btheta)\pi(\btheta|\y_{s(di)}) \,d
\btheta.
\]
Using similar arguments as those in Appendix~\ref{Appendix3}, it is easy to see
that the CPO can be obtained as
\[
\mathrm{CPO}_{di}= \biggl\{\int\frac{\pi(\btheta|\y
_{s(di)})}{f(Y_{di}|\btheta)} \,d\btheta\biggr
\}^{-1}\approx\Biggl\{\frac{1}{H}\sum
_{h=1}^H \frac
{1}{f(Y_{di}|\btheta^{(h)})} \Biggr\}^{-1}.
\]
Small values of $\mathrm{CPO}_{di}$ point out to observations that
are surprising in light of the knowledge of the other
observations [\citet{Pet90}; \citet{Ntz09}].

%s5 #&#
\section{Simulation study}\label{FreVal}
The great social relevance of poverty estimation obliges us to use methods
that are widely accepted beyond the Bayesian
community. The EB estimators introduced in \citet{MolRao10}
are highly efficient (approximately the ``best'' according to mean
squared error) under the assumed \mbox{frequentist} model. It raises the
question whether the HB procedure introduced in Section~\ref{HBmethod}
also offers good frequentist properties. To answer this question, a
simulation experiment
was conducted under the frequentist setup. In this simulation study, HB
\mbox{estimators} of poverty incidence and gap are compared
with the alternative EB estimators of \citet{MolRao10}. For this, unit
level data were generated similarly as in \citet{MolRao10}. The
population was composed of $N={}$20,000 units
distributed in $D=80$ areas with $N_d=250$ units in each area,
$d=1,\ldots,D$. Imitating a situation in which only categorical
auxiliary variables are available, as in the application with Spanish
data described in Section~\ref{applic},
we considered two dummies $X_1$ and $X_2$ as explanatory variables in
the model, apart from the intercept.
The population values of these variables were generated as $X_k\sim
\operatorname{Bin}(1,p_{kd})$, $k=1,2$, with
success probabilities $p_{1d}=0.3+0.5 d/D$ and
$p_{2d}=0.2$,
$d=1,\ldots,D$, and held fixed. We took $\bolds{\beta
}=(3,0.03,-0.04)'$, $\sigma^2=0.5^2$ and
$\rho=0.82$, so that $\sigma_u^2=0.15^2$ as in \citet{MolRao10}.
Then, using the population values of the auxiliary
variables, population responses were generated from (\ref
{condit})--(\ref{ud}) with $w_{di}=1$ for all $i$ and $d$.
The poverty line is taken as $z=12$. This value is roughly equal to
0.6 times the median welfare for
a population generated as described before, which is the
official poverty line used in EU countries. With this poverty line,
the population
poverty incidence is about 16\%. A sample $s_d$ of size $n_d=50$
is drawn by simple random sampling without replacement from area~$d$,
for $d=1,\ldots,D$, independently for all
areas. Let $s=\bigcup_{d=1}^D s_d$ be the whole sample and let $(\y
_s,\X_s)$ be the sample data.

For a given population and sample generated as described above, HB
estimates were computed as
follows. Generate $H=1000$ independent samples from the posterior
predictive distribution
of $F_{\alpha d}$ by implementing the following steps, where new
notation is defined in Appendix~\ref{posterior}:
\begin{longlist}[(3)]
\item[(1)] \textit{Generation of intra-class correlation coefficient}
$\rho^{(h)}$: Take a grid of
\mbox{$R=1000$} points in the interval $[\varepsilon,1-\varepsilon]$, for
$\varepsilon=0.0005$,
\[
\rho_r=(r-0.5)/R,\qquad r=1,\ldots,R-1.
\]
Let us define the kernel of the posterior
density of $\rho$ as
\[
k_4(\rho)= \biggl(\frac{1-\rho}{\rho} \biggr)^{D/2}\bigl|
\mathbf{Q}(\rho)\bigr|^{-1/2}\gamma(\rho)^{-(n-p)/2}\prod
_{d=1}^D \lambda_d^{1/2}(
\rho).
\]
Calculate $k_4(\rho_r)$, $r=1,2,\ldots,R-1$ and take
\[
\pi_4(\rho_r)=\frac{k_4(\rho_r)}{\sum_{r=1}^R k_4(\rho_r)},\qquad
r=1,2,\ldots,R-1.
\]
Then generate $\rho^{(h)}$ from the discrete distribution
$\{\rho_r,\pi_4(\rho_r)\}_{r=1}^{R-1}$.
Since these generated values are discrete, jitter each generated
value by adding to it a uniform random number in the interval
$(0,1/R)$.
\item[(2)] \textit{Generation of error variance}: First draw
$\sigma^{-2(h)}$ from the distribution
\[
\sigma^{-2(h)}|\rho,\y_s \sim\operatorname{Gamma} \biggl(
\frac{n-p}{2}, \frac{\gamma^{(h)}}{2} \biggr),
\]
where $\gamma^{(h)}=\gamma(\rho^{(h)})$. Then, take
$\sigma^{2(h)}=1/\sigma^{-2(h)}$.
\item[(3)] \textit{Generation of regression coefficients}: Draw $\bolds
{\beta}^{(h)}$ from the distribution
\[
\bolds{\beta}^{(h)}|\sigma^{2(h)},\rho^{(h)},
\y_s\sim N \bigl(\hat{\bolds{\beta}}^{(h)},
\sigma^{2(h)}\bigl(\mathbf{Q}^{(h)}\bigr)^{-1} \bigr),
\]
where $\hat{\bolds{\beta}}^{(h)}=\hat{\bolds{\beta}}(\rho^{(h)})$ and
$\mathbf{Q}^{(h)}=\mathbf{Q}(\rho^{(h)})$.
\item[(4)] \textit{Generation of random area effects}:
Draw $\{u_d^{(h)};d=1,\ldots,D\}$ from
\[
u_d^{(h)}|\bolds{\beta}^{(h)},
\sigma^{2(h)},\rho^{(h)},\y_s \stackrel{\mathrm{ind}}\sim N
\biggl[\lambda_d^{(h)}\bigl(\bar y_d-\bar
\x_d'\bolds{\beta}^{(h)}\bigr),\bigl(1-
\lambda_d^{(h)}\bigr)\frac{\sigma
^{2(h)}\rho^{(h)}}{1-\rho^{(h)}} \biggr],
\]
where $\lambda_d^{(h)}=\lambda_d(\rho^{(h)})$, $d=1,\ldots,D$.
\item[(5)] \textit{Generation of out-of-sample elements}:
Draw $Y_{di}^{(h)}$, $i\in r_d$, from their
distribution given all parameters
$\btheta^{(h)}=(u_1^{(h)},\ldots,u_D^{(h)},\bolds{\beta}^{(h)},\sigma
^{2(h)}\rho^{(h)})'$,
given by
\[
Y_{di}^{(h)}|\y_s,\btheta^{(h)}\sim N
\bigl(\x_{di}'\bolds{\beta}^{(h)}+u_d^{(h)},
\sigma^{2(h)}\bigr),\qquad i\in r_d.
\]
Then $\y_{rd}^{(h)}=\{Y_{di}^{(h)}; i\in r_d\}$ is the vector
containing all generated out-of-sample elements from domain $d$,
$d=1,\ldots,D$.
\item[(6)] \textit{Calculation of poverty indicator}: Consider the
vector with
sample elements attached to generated out-of-sample elements from
domain $d$,
$\y_d^{(h)}=(\y_{sd}',(\y_{rd}^{(h)})')'$. Calculate
the poverty indicator for domain $d$ as in (\ref{MCF}) using $\y_d^{(h)}$,
$d=1,\ldots,D$.
\end{longlist}
At the end of steps (1)--(6), we get a sample of
independent values $F_{\alpha d}^{(h)}$, $h=1,\ldots,H$. Finally,
compute the posterior mean and the posterior variance as indicated in
(\ref{MCest}).

A total of $I=1000$ population vectors $\y^{(i)}$ were
generated from the true model described above. For each population
$i=1,\ldots,I$, the following process was
repeated: first, we calculated true area poverty incidences and gaps;
then, we selected the population elements corresponding to
the sample indices, assuming that those sample indices are constant
over Monte Carlo
simulations, that is, following a strictly model-based approach. Using
the sample elements, we computed EB
estimates following the procedure in \citet{MolRao10}, and HB
estimates using the approach described in (1)--(6).

%
%f1 #&#
\begin{figure}%[h]

\includegraphics{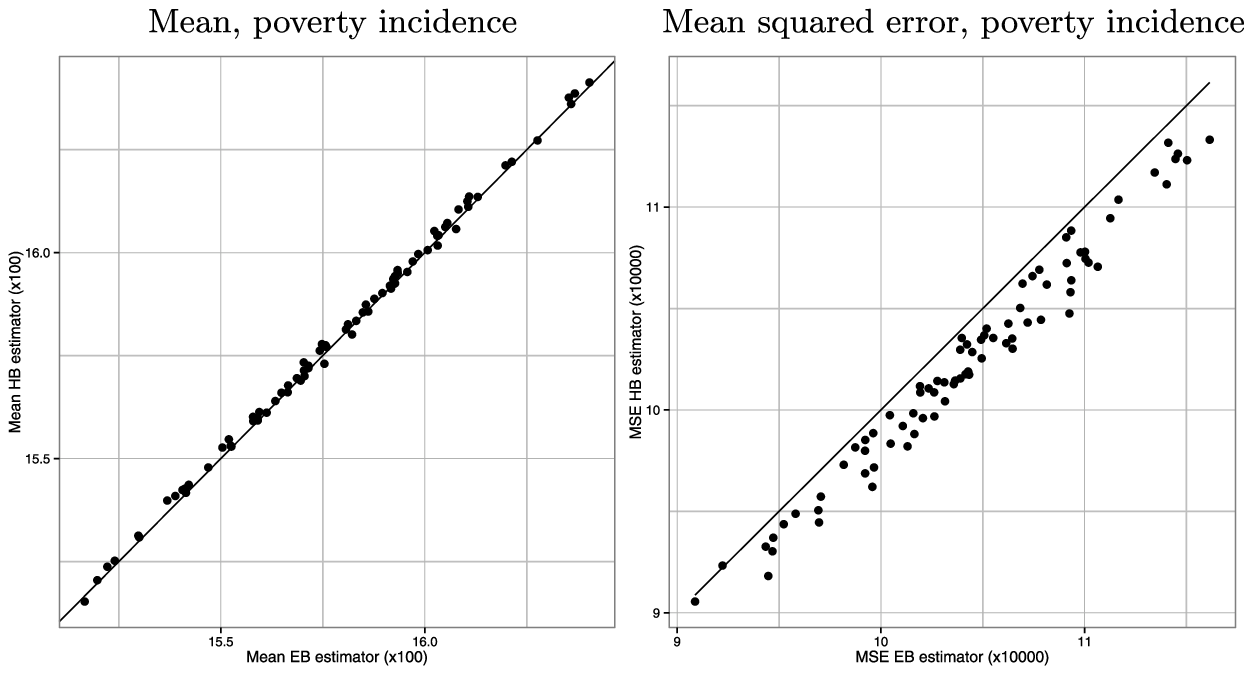}

\caption{On the left, means $\times$100 over
simulated populations of
HB estimates of poverty incidence $F_{0 d}$ against the analogous
means for EB estimates, for each area $d$.
On the right, mean squared errors ${\times}$10$^4$ of HB estimators
against those of EB estimators.}\label{plotDispPI}
\end{figure}

\begin{figure}[t]

\includegraphics{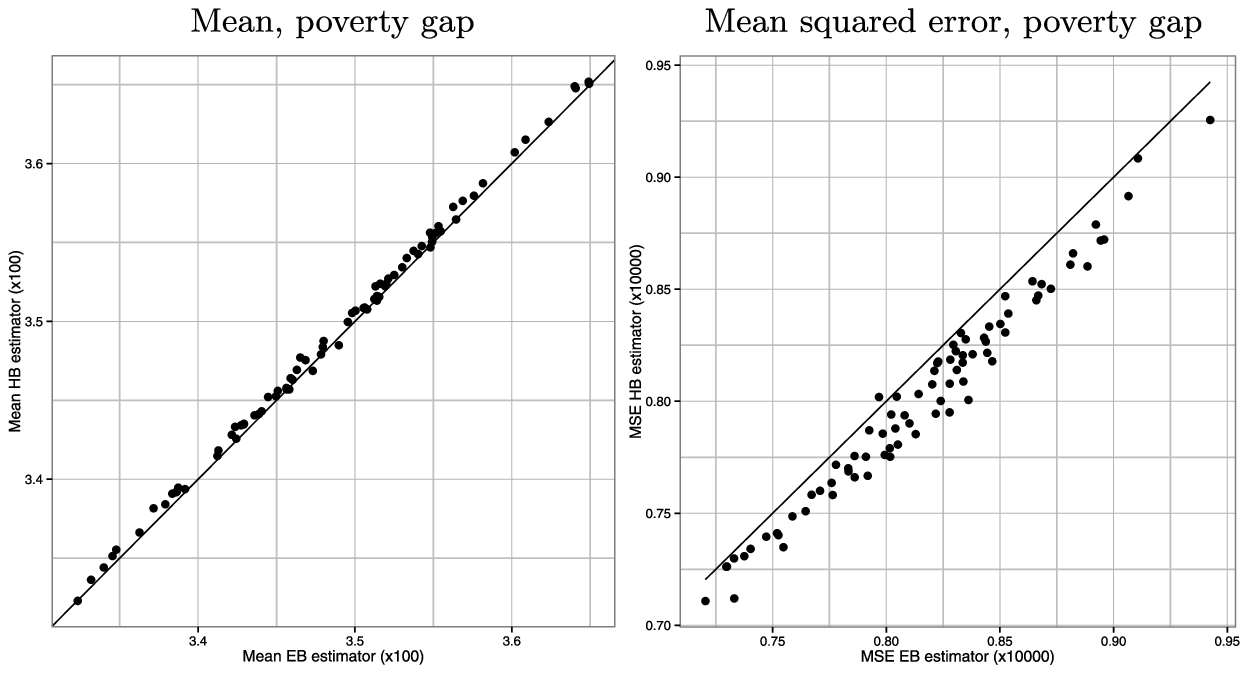}

\caption{On the left, means $\times$100 over
simulated populations of
HB estimators of the poverty gap $F_{1 d}$ against the analogous means
for EB estimators, for each area $d$.
On the right, mean squared errors ${\times}$10$^4$ of HB estimators
against those of EB estimators.}\label{plotDispPG}
\end{figure}

The left panel of Figure~\ref{plotDispPI} displays means over Monte
Carlo replicates
of HB estimates of poverty incidences for each
area against corresponding means of EB estimates. The right panel
gives the frequentist mean squared
errors of HB\vadjust{\goodbreak} estimates against those of EB estimates for each area.
Thus, from a frequentist point of view, we can see that the two
estimators are practically the same,
probably because only noninformative priors have been considered.
The true mean squared errors of EB estimators are slightly smaller
than those of HB estimates, which is somewhat
sensible since the EB estimates are approximately the best under the
frequentist paradigm.
Figure~\ref{plotDispPG} shows similar results for the poverty gap.
Both figures show that the
HB estimates display good frequentist properties.

\begin{figure}[b]

\includegraphics{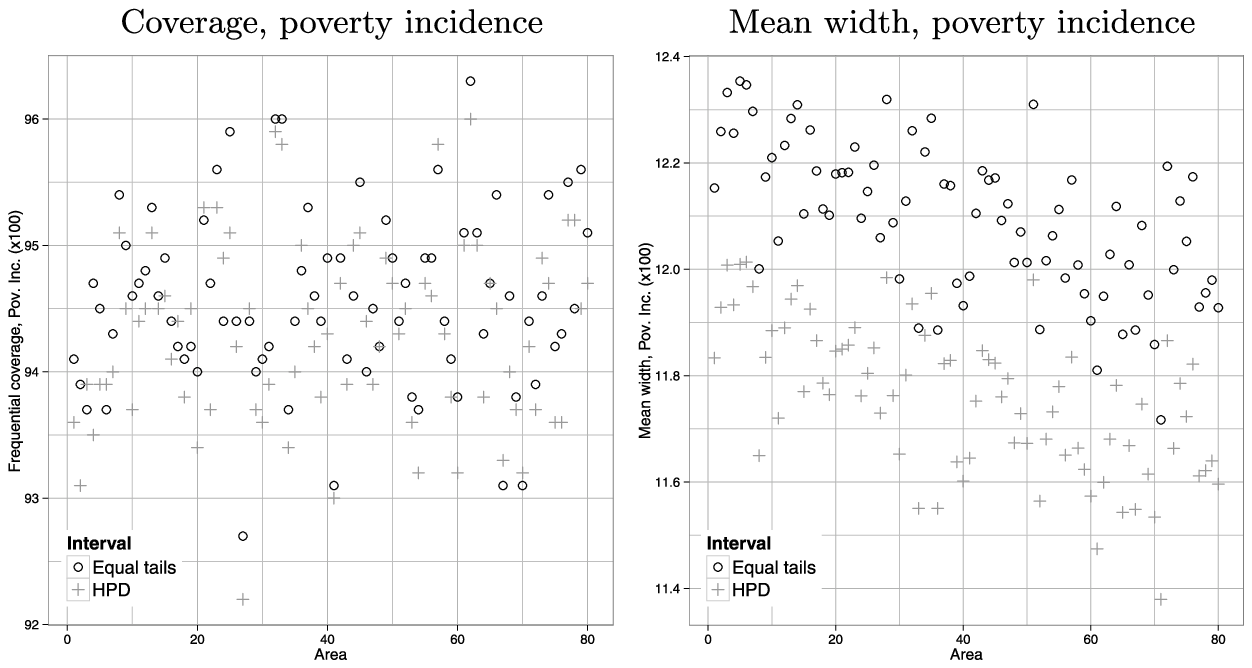}

\caption{Percent coverage, left panel, and mean
widths, right panel, over Monte Carlo populations of
equal tails and highest posterior density intervals for poverty
incidence $F_{0 d}$ for each area $d$.}\label{plotCoverPI}
\end{figure}

\begin{figure}[t]

\includegraphics{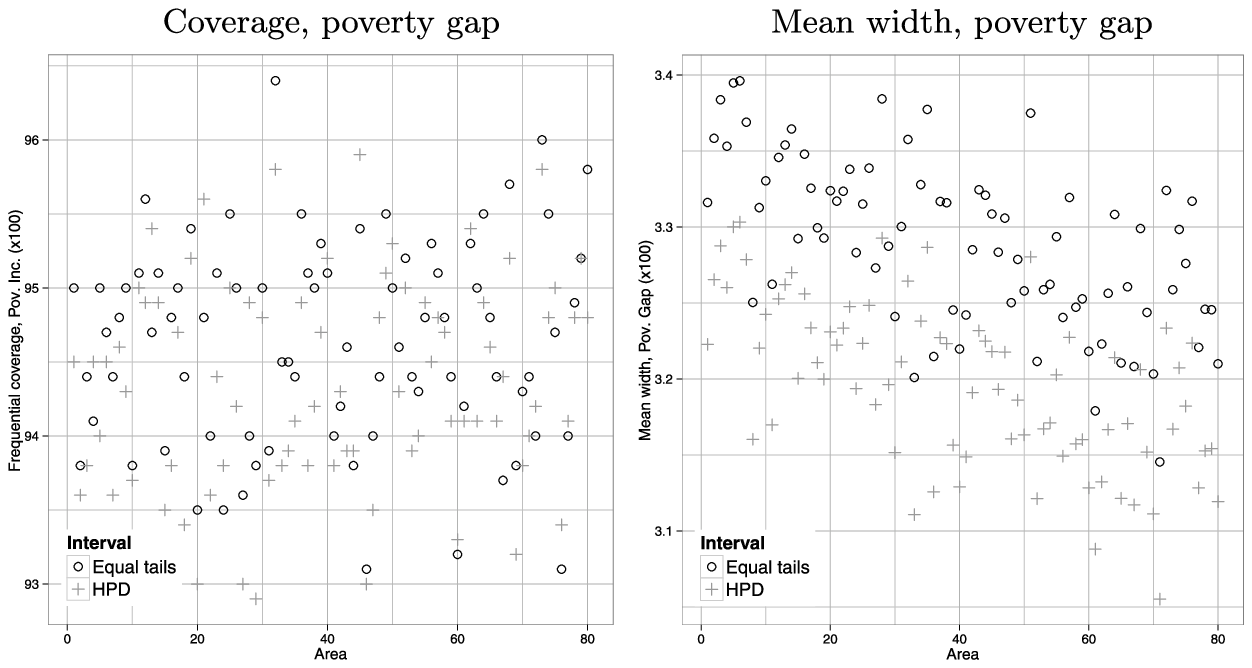}

\caption{Percent coverage, left panel, and mean
widths, right panel, over Monte Carlo populations of
equal tails and highest posterior density credible intervals for
poverty gap $F_{1 d}$ for each area $d$.}\label{plotCoverPG}
\end{figure}

In addition to point estimates, the HB approach can also deliver
credible intervals. It is interesting to see whether these intervals
satisfy the basic frequentist property of covering the true value. The
left panel of Figure~\ref{plotCoverPI} displays the frequentist
coverage of 95\%
credible intervals for the area poverty incidences, calculated as
a percentage of Monte Carlo replicates in which credible intervals
contain true
values. We have plotted the coverages of equal tails credible
intervals together
with those of highest probability density intervals [\citet{CheSha99}].
This figure reveals a slight undercoverage of less than 1\% for the
two types of intervals. The estimated
coverage of credible intervals with only $H=1000$ replicates might not
be very accurate, so we
guess that a larger $H$ could show a smaller undercoverage. On the
right panel of the same figure
we report the mean widths of the two types of intervals. As expected,
the highest posterior density intervals are clearly narrower.
Similar conclusions can be drawn from Figure~\ref{plotCoverPG} for the
poverty gap.

%
%f2 #&#

%
%f3 #&#

%
%f4 #&#

%
%f5 #&#
\begin{figure}[b]

\includegraphics{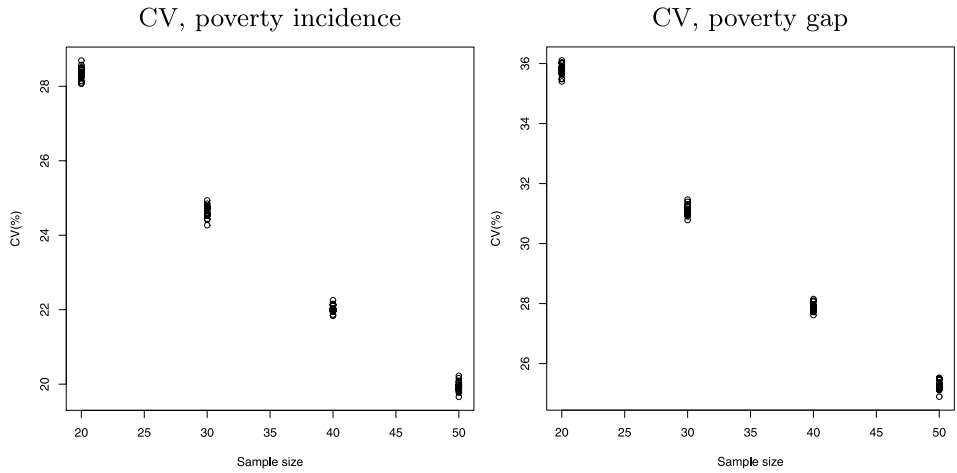}

\caption{Mean over Monte Carlo simulations of CVs
of HB estimators of poverty incidence (left panel) and poverty gap
(right panel).}\label{plotCVPIPG}
\end{figure}

Finally, to analyze the effect of the sample size on the
performance of the HB estimators, a new simulation experiment was
conducted with increasing area sample sizes $n_d$ in the set $\{20,
30, 40, 50\}$,
with each value repeated for 20 areas and with a total number of areas
$D=20\times4=80$ as before. In this experiment we
omitted the covariates that could distort the results and
consider only a mean model with intercept $\bolds{\beta}_0=3$ as
before. Figure~\ref{plotCVPIPG}
plots the mean coefficients of variation (CVs) of HB estimators of
poverty incidences
and poverty gaps. The estimated CV of an HB estimator is taken as
the square root of the posterior variance divided by the estimate.
Observe that, on average, the CVs increase about 3\%
when decreasing the area sample size in 10 units. Moreover, in
this simulated example, it turns out that at least $n_d=50$ units need
to be observed in area $d$
to keep the CV of HB estimators of poverty incidences below
20\%. For the poverty gap, which is more difficult to estimate,
the same sample size ensures a maximum CV of~25\%.

%s6 #&#
\section{Poverty mapping in Spain}\label{applic}
This section describes an application of the proposed HB method
to poverty mapping in Spanish provinces by gender.
The data come from the SILC conducted in Spain in year
2006 and is the same used by \citet{MolRao10}. The SILC collects
microdata on income, poverty, social exclusion and living conditions,
in a
timely and comparable way across European Union (EU) countries.
The results of this survey are then used for the structural indicators
of social cohesion such as poverty incidence, income quintile share
ratio and gender pay
gap. Indeed, equality between women and men is one of the EU's
founding values;
see, for example, \url{http://ec.europa.eu/justice/gender-equality/}. Thus,
the EU is especially
concerned about gender issues, fostering research devoted to the
quantification or measurement of equality.
For example, one of the commitments of the SAMPLE project funded by
the European Commission (\url{http://www.sample-project.eu/}) was to obtain
poverty indicators in Spanish provinces by gender.

The Spanish SILC survey design is as follows. An independent sample is
drawn from each of the Spanish Autonomous
Communities using a two-stage design with stratification of the
first stage units. The first stage or primary sampling units are
census tracks and they are grouped into strata
according to the size of the municipality where the census track is located.
Census tracks are drawn within each stratum with probability
proportional to their size.
The secondary sampling units are main family dwellings, which are
selected with equal probability and with random start systematic sampling.
Within those last stage units, all individuals with usual residence in
the dwelling are interviewed.
This procedure results in self-weighted samples within each stratum.
This survey is planned to provide reliable estimates only for the
overall Spain and for the
Autonomous Communities which are large Spanish regions, but it cannot deliver
efficient estimates for the Spanish provinces disaggregated by gender
due to the small sample
size (provinces are nested within Autonomous Communities). Therefore,
small area estimation techniques that ``borrow strength'' from other
provinces are
needed. The HB methodology proposed in this paper allows us to
produce efficient estimates of practically any poverty indicator
for the Spanish provinces by gender, using a computationally fast
procedure, and provides at the same time all pertinent output such as
uncertainty
measures and credible intervals.\looseness=-1

In this application, the target domains are the $D=52$ Spanish provinces.
Since many studies on poverty in developing countries point to more
severe levels of poverty
for females than for males, it is very important to analyze if this
happens in
Spain as well. Thus, we are interested in giving estimates also by
gender. To this end, we applied the HB procedure
described in Section~\ref{FreVal} separately for each gender,
obtaining estimates of poverty incidences and gaps for the Spanish provinces,
together with 95\% highest posterior density intervals. The HB
procedure was applied with
a grid of $R=1000$ values of $\rho$ and $H=1000$ Monte Carlo replicates.
For comparison, the EB method of \citet{MolRao10} was also applied
separately for each gender. Since this method is computationally slower,
we considered only $L=50$ Monte Carlo simulations as in \citet{MolRao10}.
The parametric bootstrap approach proposed in the same paper for mean
squared error (MSE)
estimation of EB estimates was applied with $B=200$ bootstrap
replicates.

The overall sample size is 16,650 for
males and 17,739 for females. The population size is 21,285,431 for males
and 21,876,953 for females, with a total population size of over 43
million. We considered the same auxiliary
variables as in \citet{MolRao10}, namely, the indicators of
five quinquennial age groups, of having Spanish nationality, of
the three levels of the variable education level, and of the three
categories of the variable labor force status, ``unemployed,''
``employed'' and ``inactive.'' For each auxiliary variable, one of
the categories was considered as base reference, omitting the
corresponding indicator and including an intercept in the model.

When making use of continuous covariates, all the methods (HB, EB
and WB) require a full census of those covariates. In this
application, however, only dummy indicators were included in the
model and, therefore, only the counts of people with the same vector
of $x$-values are needed. However, to make the computations as
general as possible, we imitated the case of having continuous
covariates by constructing the full census matrices
$\X_d=(\x_{d1},\ldots,\x_{dN_d})'$. This was done using the data
from the Spanish Labour Force Survey (LFS), which has a much
larger sample size than the SILC (155,333 as compared with 34,389)
and therefore offers information with much better quality. Each
LFS vector $\x_{di}'$ was replicated a number of times equal to
its corresponding LFS sampling weight; the resulting matrix $\X_d$
may be treated as a proxy of the true census matrix. As noted in
Section~\ref{intro}, the WB was able to secure true census matrices
$\X_d$ from statistical offices of many countries.

The welfare variable provided by the SILC for each individual and
used to measure poverty by the Spanish Statistical
Institute (INE in Spanish) and also by the European Statistical Office
Eurostat is the so-called
equivalized annual net income, which is the household annual net income,
divided by a measure of household size calculated according to the
scale defined by the OCDE.
The resulting quantity can be interpreted as a kind of per
capita income and for this reason it is assigned to each household member.
Using instead the total household income would require the definition of
a different poverty line for each possible household size and would
not allow us to estimate by gender. For this reason, in this application
we consider that the units are the individuals and, as welfare measure $E_{di}$,
we consider the equivalized annual net income. Due to the clear right skewness
of the histogram of $E_{di}$ values, we
consider the transformation $Y_{di}=T(E_{di})=\log(E_{di}+c)$,
where $c\geq\max\{0,-\min(E_{dj})+1\}$ is a constant
selected in such a way that all shifted incomes $E_{di}+c$ are
positive (there are few negative $E_{di}$) and for which the
distribution of model residuals is closest to being symmetric. To
select $c$, we took a grid of points in the range of income values and
the model was
fitted for each point in the grid. Then $c$ was selected as the point
in this grid
for which Fisher's asymmetry coefficient of model residuals (third
order centered sample moment divided by the cube standard deviation)
was closest to zero. It turned out to be exactly the same value in the
two models for Males and Females.

%
%f6 #&#
\begin{figure}

\includegraphics{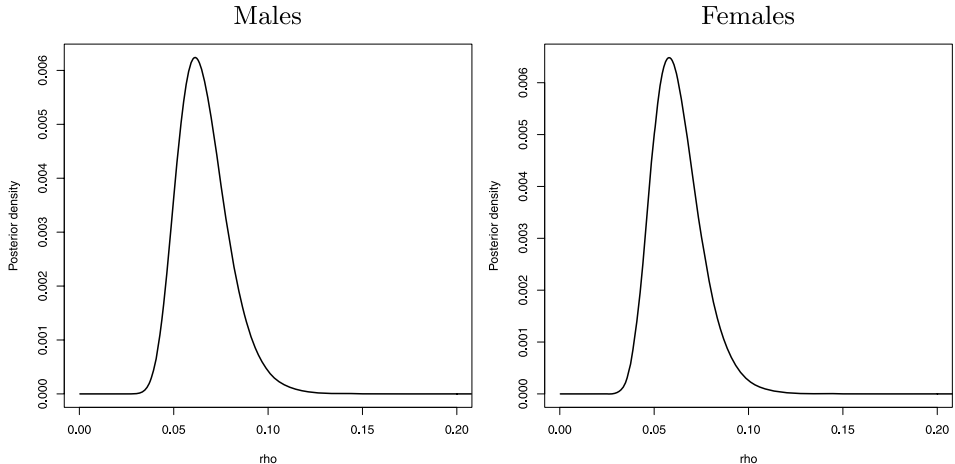}

\caption{Posterior density of $\rho$ in the model for males, left
panel, and
females, right panel, obtained
drawing from a grid in $[\varepsilon,1-\varepsilon]$ with $\varepsilon
=0.0001$.}\label{PostRhoMales}
\end{figure}

%
%f7 #&#
\begin{figure}

\includegraphics{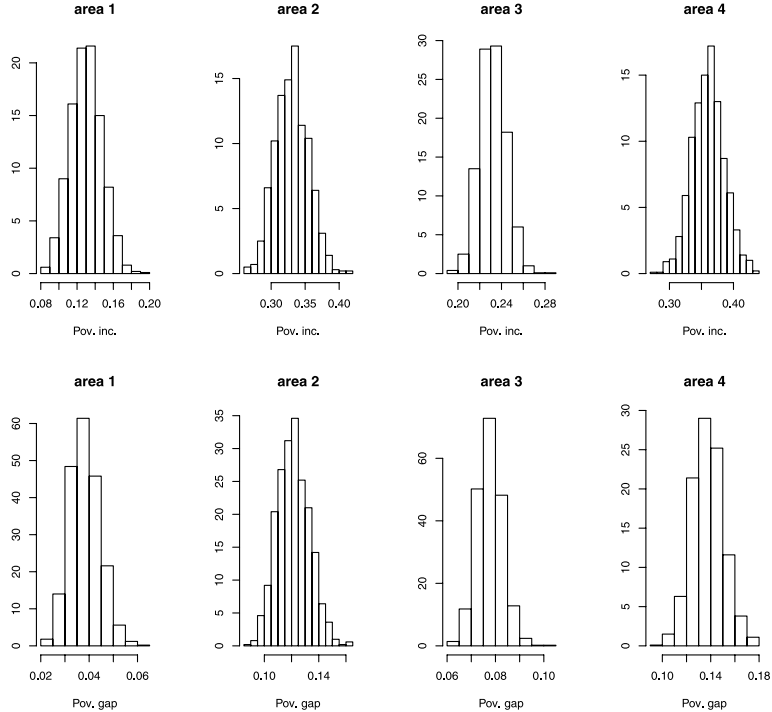}

\caption{Histograms of posterior distributions of poverty incidences,
upper panel, and poverty gaps, lower panel,
for areas $d=1,2,3,4$, respectively, in columns, for females.}\label{HistPostDist}
\end{figure}

Since samples are drawn independently for each of the 18
Autonomous Communities and these regions might have different
socio-economic levels,
trying to accommodate the sampling design, we also fitted a model with
Autonomous Community effects.
However, the goodness of fit of the model including these effects, as
measured by AIC and BIC,
became worse for the two genders. %In a model with province random
%effects nested within Autonomous Communities
%random effects, the goodness-of-fit measures got a very mild
%improvement
%(BIC=11707.24 versus BIC=11703.12 for Males, and BIC=15009.38 versus
%BIC=15003.2 for females).
%This indicates that province random effects are explaining to a large
%extent the Autonomous Community effects.
%Note that several Autonomous Communities are composed of only one
%province.
Thus, we consider a more parsimonious model without Autonomous
Community effects.

Figure~\ref{PostRhoMales} shows the posterior density of the intraclass
correlation $\rho$ obtained in the models for males and females, using
a grid of 5000 values
of $\rho$ in the interval $[0.0001,0.9999]$. The two plots show that
the mass of the
posterior density of $\rho$ is mostly concentrated in
$[0.04,0.1]$ and, therefore, the use of the truncation point $\varepsilon
=0.0001$ in the two
extremes of the range of $\rho$ to ensure a proper posterior
does not have any effect in this application. In fact, trying to
analyze the sensitivity to varying $\varepsilon$,
we also used $\varepsilon=0.001$ and $\varepsilon=0.005$ and we found
virtually no difference in the resulting posterior densities of $\rho$.

%
%f8 #&#
\begin{figure}%[p]

\includegraphics{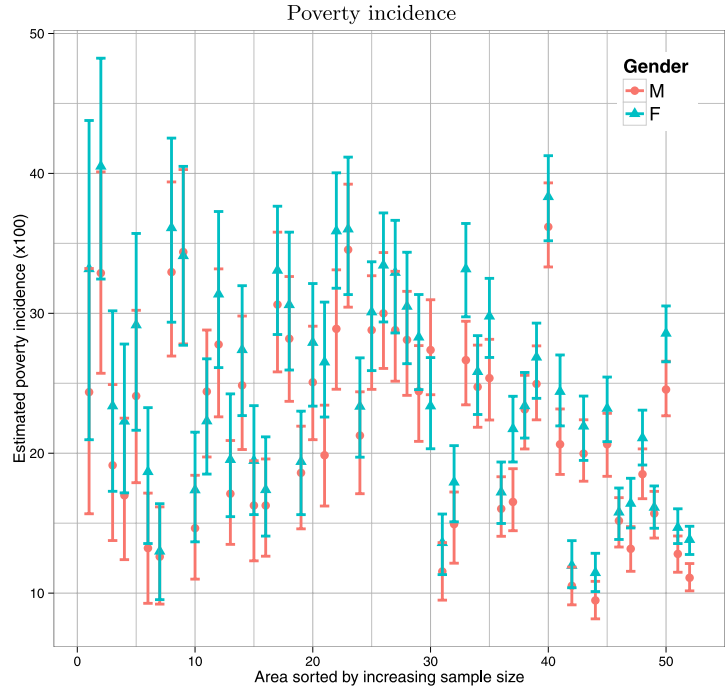}

\caption{Hierarchical Bayes estimates of
poverty incidences
with highest posterior density intervals for each gender and for each
area $d$. Areas are sorted by increasing sample size.}\label{plotHPDintprop}
\end{figure}

%f9 #&#
\begin{figure}%[h]

\includegraphics{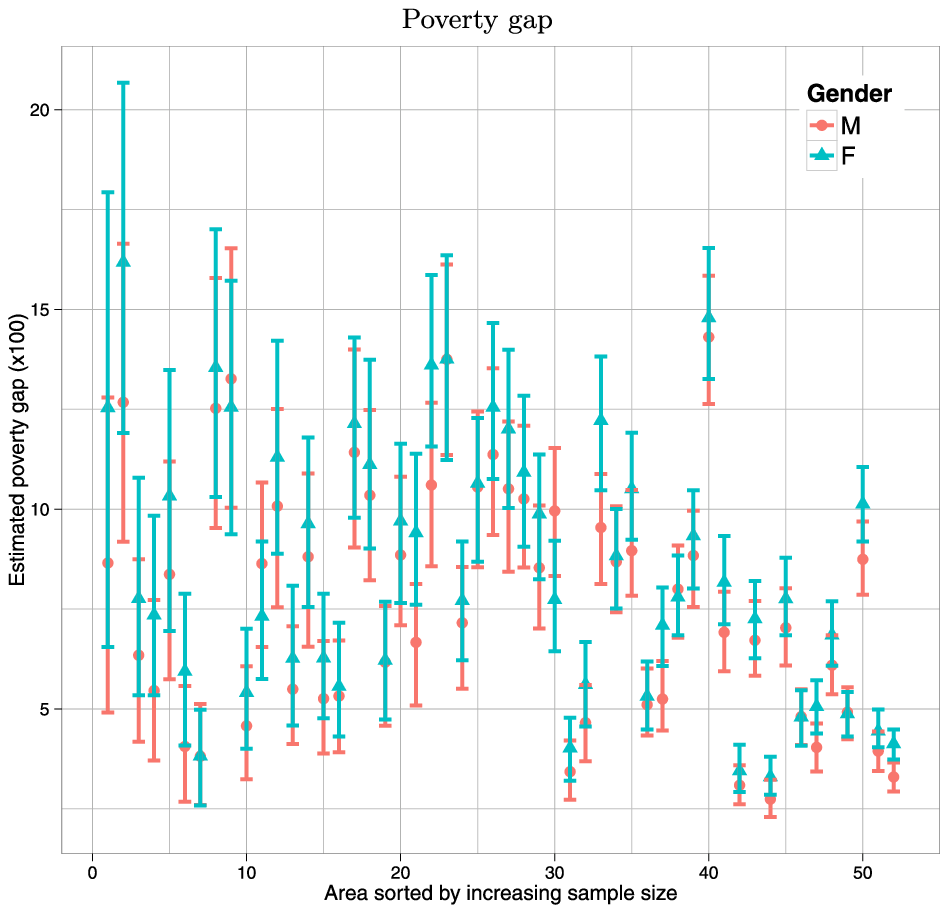}

\caption{Hierarchical Bayes estimates of poverty
gaps with highest posterior density
intervals for each gender and for each area $d$. Areas are sorted by
increasing sample size.}\label{plotHPDintgap}
\end{figure}

%
%f10 #&#
\begin{figure}%[h]

\includegraphics{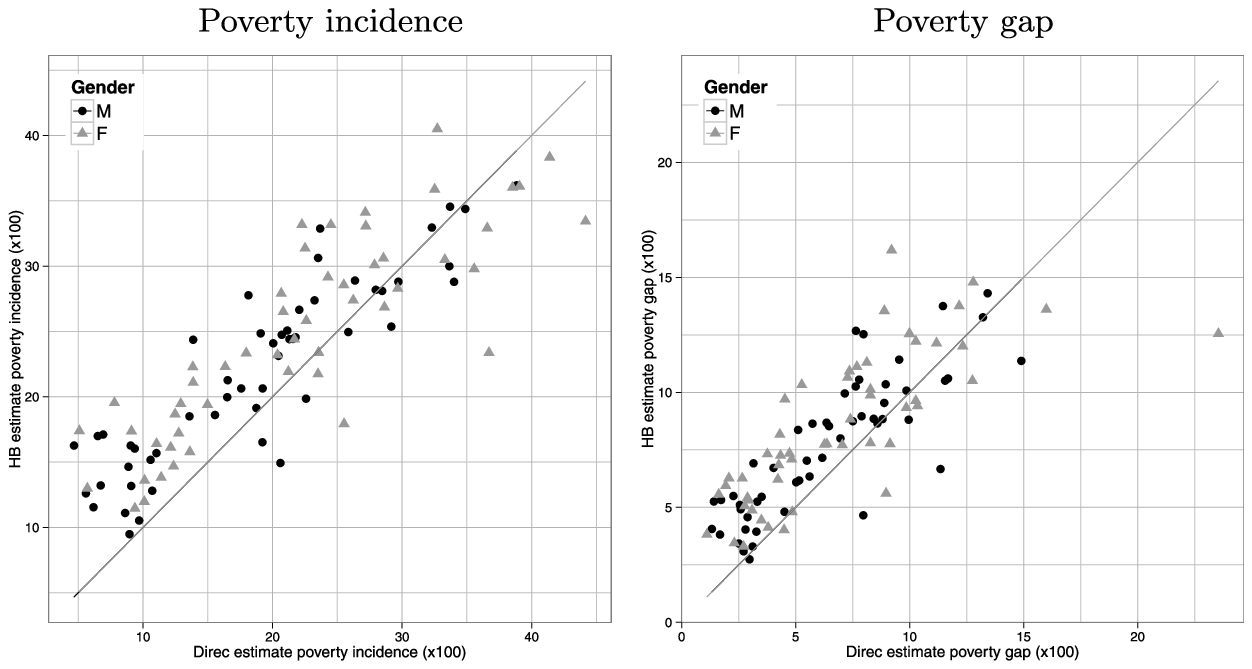}

\caption{Hierarchical Bayes estimates of poverty incidence $F_{0 d}$, left
panel, and of poverty gap $F_{1 d}$, right panel,
against direct estimates for each province $d$.}\label{plotDispHBdir}
\end{figure}

Figure~\ref{HistPostDist} shows the histograms of the posterior
distributions of poverty incidences, upper panel, and poverty gaps,
lower panel, for the first 4
provinces in the alphabetical order, in the model for females. Note
that these histograms are slightly skewed. Therefore,
we considered highest posterior density intervals instead of
credible intervals with equal probability tails. These intervals
were computed as described in \citet{CheSha99}. Figure~\ref{plotHPDintprop}
plots HB estimates of poverty incidences together with their
corresponding 95\%
highest posterior density intervals for each gender and for each
province, with
areas sorted by increasing sample size. This figure shows that the
length of the intervals decreases
as the area sample sizes increase, as expected. It also shows that the
estimated poverty incidences for males are smaller than for females
for most provinces, although the two corresponding intervals cross
each other for practically all provinces.
Concerning the poverty gap, which measures the degree of poverty
instead of the frequency of poor,
Figure~\ref{plotHPDintgap} shows a very similar pattern, with point
estimates for females larger than for males in most provinces.

%f11 #&#
\begin{figure}[t]

\includegraphics{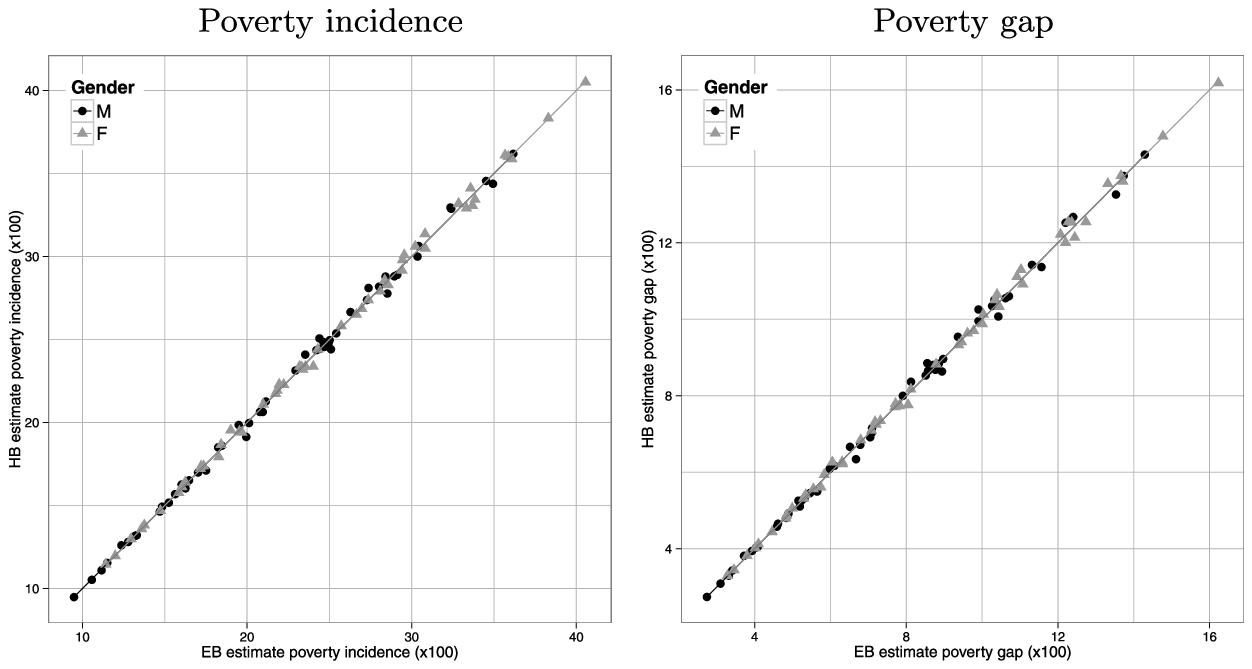}

\caption{Hierarchical Bayes estimates of poverty incidence $F_{0 d}$, left
panel, and of poverty gap $F_{1 d}$, right panel,
against EB estimates for each province $d$.}\label{plotDispHBEB}
\end{figure}

%
%f12 #&#
\begin{figure}[b]

\includegraphics{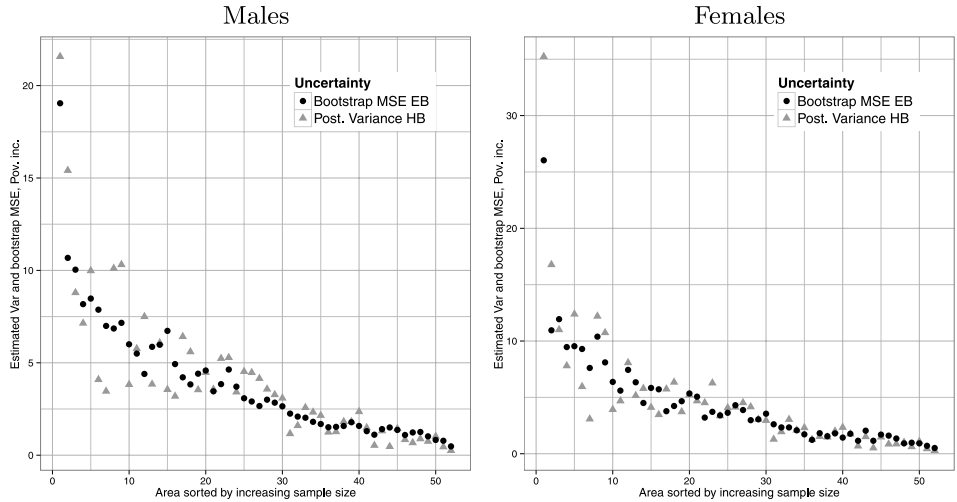}

\caption{Posterior variances of HB estimators and bootstrap
mean squared errors of EB estimators of poverty incidence for each
province for males,
left panel, and females, right panel. Provinces sorted by increasing
sample size.}\label{ErrorMeasure}
\end{figure}

%f13 #&#
\begin{figure}

\includegraphics{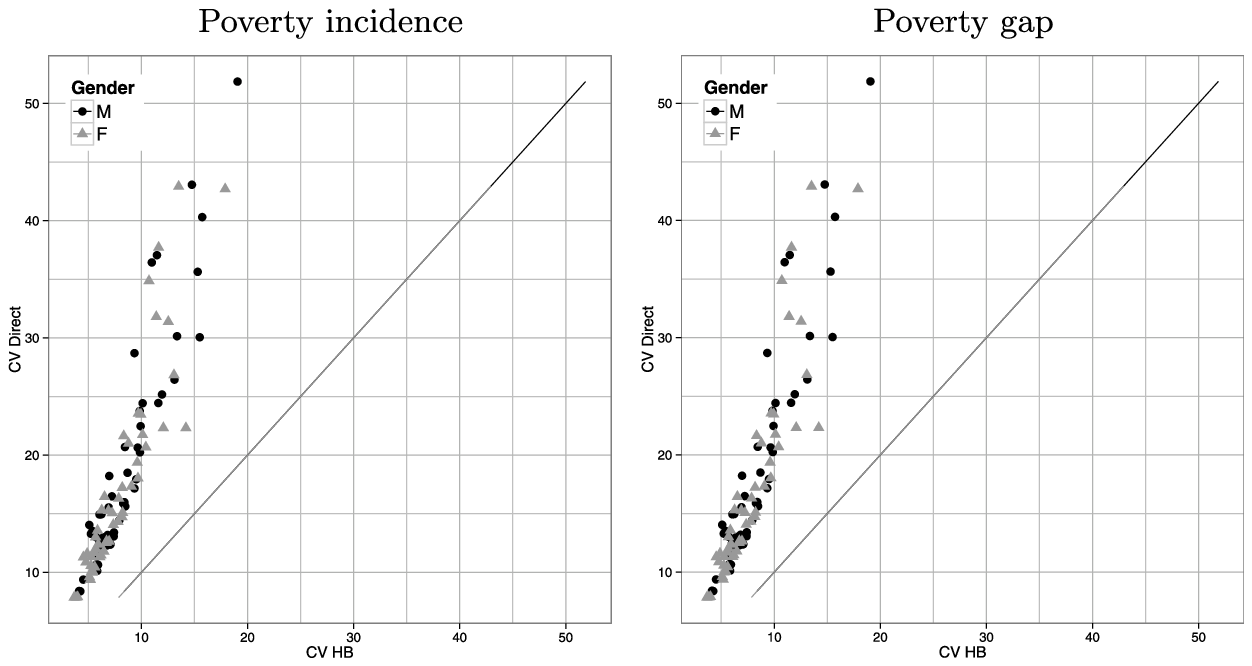}

\caption{Coefficients of variation of direct estimates of poverty incidence
$F_{0 d}$, left panel,
and of poverty gap $F_{1 d}$, right panel, against coefficients of
variation of HB estimates for each province $d$.}\label{plotCVDirecHB}
\end{figure}

Figure~\ref{plotDispHBdir} plots HB estimates of poverty incidence,
left panel, and of poverty gap,
right panel, against direct estimates for each area $d$, using
separate plotting symbols for each gender.
Observe that all points lie around the line, except for one of them
corresponding to the poverty gap for
women. This point is separated from the line because its direct
estimate is much larger than the HB estimate. This occurs because
HB estimates shrink extreme direct estimates toward synthetic
regression ones for areas with small sample size.

Figure~\ref{plotDispHBEB} plots HB estimates of poverty incidence,
left panel, and of poverty gap,
right panel, against EB estimates for each gender and for each
province. Observe that HB estimates are practically
equal to the corresponding EB estimates. Thus, in this application the
point estimates obtained by the HB method
proposed in this paper agree to a great extent with those obtained by
the EB
method.

Turning to the measures of variability of EB and HB estimators,
Figure~\ref{ErrorMeasure}
compares the estimated MSEs of the EB estimators obtained by the
parametric bootstrap described in \citet{MolRao10},
with the posterior variances. Although in principle these measures are
not strictly comparable, it is interesting to see their
similarity, and this similarity increases for areas with larger sample
sizes.

Concerning computational efficiency, in this application, the full EB procedure
consisting in the Monte Carlo approximation of the EB estimator with
$L=50$ Monte Carlo replicates
and the bootstrap method for MSE estimation with $B=200$ bootstrap
replicates took 44.2 hours in a 2.67~GHz
PC, whereas the HB procedure takes 3.7 hours. If we wanted the
bootstrap MSE estimates to have comparable precision
as the HB posterior variances and take $B=H=1000$ bootstrap
replicates, the computational time of the full EB method
in this application would be over 9 nonstop days on the same computer.
Use of double bootstrap for bias reduction of the bootstrap MSE estimator
would increase the computational complexity manyfold. Thus, for a
larger number of auxiliary variables $p$,
larger population size or a more complex indicator, computational
times might be considerable for the EB method and in those cases the
HB method represents a much faster alternative.

To see more clearly the efficiency gain of the HB estimates over
direct estimates,
in Figure~\ref{plotCVDirecHB} we plot the estimated CVs of direct
estimators against those of
HB estimators for all provinces. Observe that the CVs of direct
estimators are above the 45$^\circ$ line for
all the provinces, indicating that HB estimates are more precise than
direct estimates for all the provinces. Moreover, the gains are larger
for provinces with smaller sample sizes and
can be considerably large for some of the provinces. In contrast, the
results obtained by \citet{MolMor09}
under an area-level model using the aggregated values of the same
covariates provided only marginal reductions in the CVs over
direct estimates.

%
%f14 #&#
\begin{figure}[t]

\includegraphics{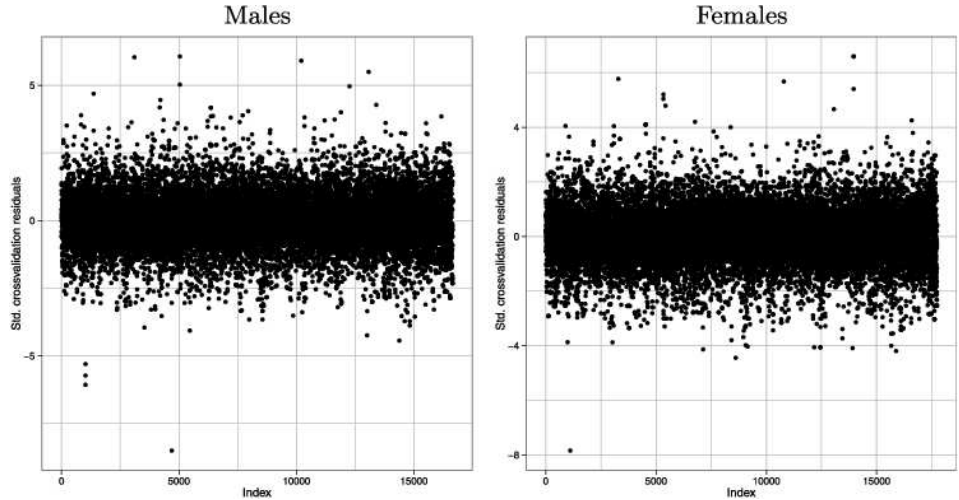}

\caption{Index plot of standardized residuals for
males, left panel, and females, right panel.}\label{StdDelRes}
\end{figure}

%f15 #&#
\begin{figure}[b]

\includegraphics{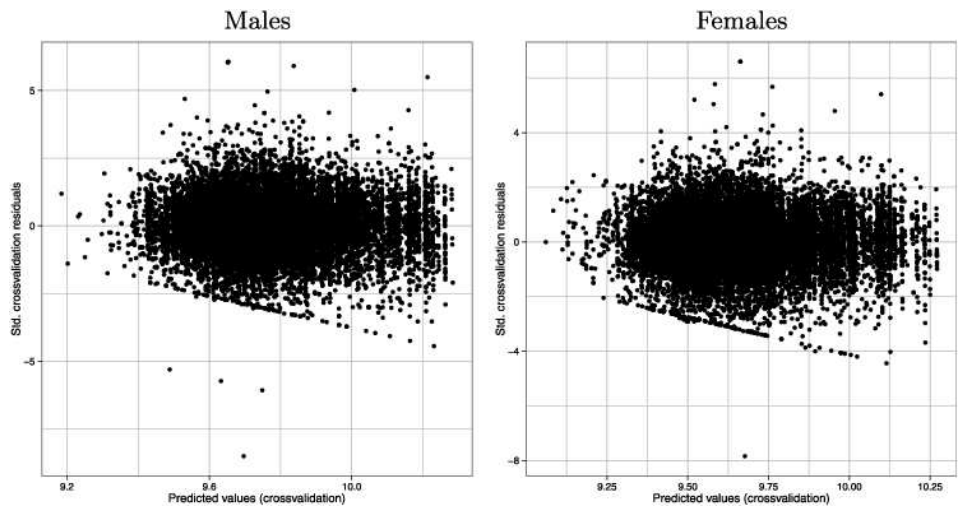}

\caption{Standardized residuals against predicted
values by cross-validation, for males, left panel, and females, right panel.}\label{StdDelResPred}
\end{figure}

\begin{figure}[t]

\includegraphics{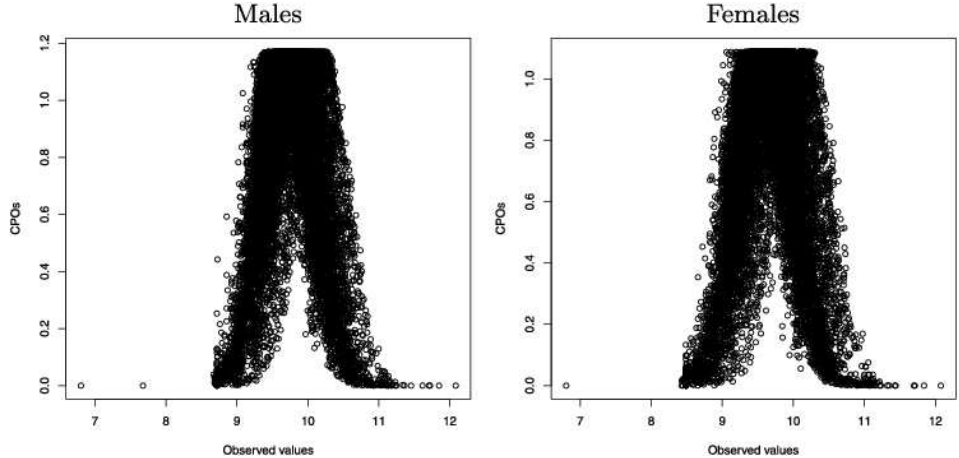}

\caption{CPOs against observed values $Y_{di}$ for
males, left panel, and females, right panel.}\label{CPOobs}
\end{figure}

We have also done some diagnostic checks of the model assumptions
using the cross-validation residuals $r_{di}$ introduced in (\ref{delres}).
Index plots of residuals for males and females are included in
Figure~\ref{StdDelRes}. In Figure~\ref{StdDelResPred} we show
the plots of standardized cross-validation residuals against predicted values.
The points that appear aligned at the bottom correspond to a number of
zero incomes. Apart from this fact, we can see
that the plots look acceptable without any visible pattern.
Concerning CPOs, Figure~\ref{CPOobs} plots these validation measures
against observed values $Y_{di}$ for males and females. As expected,
there is high predicted
power near the center of the data. The percentage of observations with
CPO values
below 0.025 turns out to be 1.5\% for females and 1.2\% for
males, and the percentage below 0.014 (extreme outliers) is
0.9\% for females and 0.8\% for males. These results
do not show any indication of serious departure from model
assumptions; see \citet{Ntz09}.

%
%f16 #&#

%f17 #&#
\begin{figure}

\includegraphics{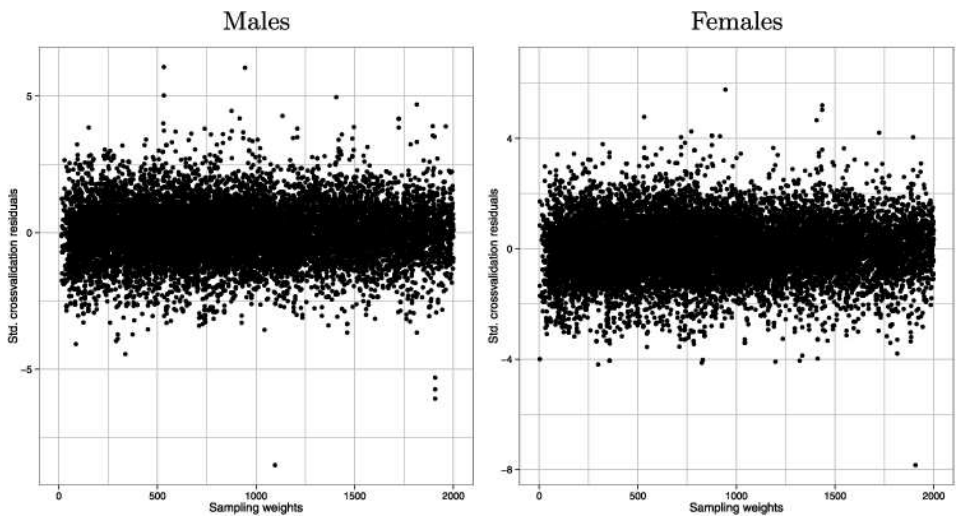}

\caption{Standardized residuals against
sampling weights in the range 0--2000, for males, left panel, and
females, right panel.}\label{ResidSamplingWeights}
\end{figure}

Note that in this method, as in any other small area estimation
procedure based on a unit-level model, the model is assumed not only
for the sample units but also for the out-of-sample units.
This assumption is reasonable as long as the design is noninformative,
that is,
the inclusion probabilities of the units in the sample are not
related to the study variable (income) after accounting for the auxiliary
variables. The SILC data contains the sampling weights or inverses of
inclusion probabilities
corrected for calibration and nonresponse. If the design is
informative, model residuals should be related somehow with sampling weights.
Thus, to analyze whether there is any evidence of informative sampling,
in Figure~\ref{ResidSamplingWeights} we have plotted cross-validation residuals
$r_{di}$ versus sampling weights for males (left) and females (right)
in the range
0--2000. There are sampling weights greater than 2000, but since
the distribution is clearly right skewed with less large weights,
for clarity of the plots we have plotted here the main part of the distribution.
The null pattern of these plots indicate no evidence of informative
sampling in this application.

Table~\ref{TablePI} reports results obtained from the estimation of
the poverty incidence,
for provinces with sample sizes closest to minimum, first quartile,
median, third quartile and maximum, for females and males.
See that the posterior coefficient of variation
is below 20\% even for the area with smallest sample size, the
province of Soria.
Table~\ref{TablePG} shows the corresponding results for the poverty
gap, where the maximum coefficient of variation is below 25\%.

%
%t1 #&#
\begin{sidewaystable}%[p]
\tablewidth=\textheight
\tablewidth=\textwidth
\caption{Sample size,
HB estimates of poverty incidence $\times$100, lower and upper
limits of highest posterior density intervals and coefficients of
variation of HB estimates for the Spanish provinces with sample
size closest to minimum, first quartile, median, third quartile
and maximum, for each gender}\label{TablePI}
\begin{tabular*}{\textwidth}{@{\extracolsep{\fill}}@{}ld{4.0}d{2.1}d{2.1}d{2.1}d{2.1}ld{4.0}d{2.1}d{2.1}d{2.1}d{2.1}@{}}
\hline
& \multicolumn{5}{c}{\textbf{Males}} & & \multicolumn{5}{c}{\textbf{Females}}\\[-6pt]
& \multicolumn{5}{c}{\hrulefill} & & \multicolumn{5}{c@{}}{\hrulefill}\\
\textbf{Province} & \multicolumn{1}{c}{$\bolds{n_d}$} & \multicolumn{1}{c}{$\bolds{\hat F_{0 d}^{\mathrm{HB}}}$} & \multicolumn{1}{c}{$\bolds{\mathrm{ll}(F_{0 d})}$}
                  & \multicolumn{1}{c}{$\bolds{\mathrm{ul}(F_{0 d})}$} & \multicolumn{1}{c}{$\bolds{\mathrm{cv}(\hat F_{0 d}^{\mathrm{HB}})}$} &
\textbf{Province} & \multicolumn{1}{c}{$\bolds{n_d}$} & \multicolumn{1}{c}{$\bolds{\hat F_{0 d}^{\mathrm{HB}}}$} & \multicolumn{1}{c}{$\bolds{\mathrm{ll}(F_{0 d})}$}
                  & \multicolumn{1}{c}{$\bolds{\mathrm{ul}(F_{0 d})}$} & \multicolumn{1}{c@{}}{$\bolds{\mathrm{cv}(\hat F_{0 d}^{\mathrm{HB}})}$} \\
\hline
Soria & 24 & 24.4 & 15.7 & 33.2 & 19.1 & Soria & 17 & 33.2 & 21.0 & 43.8 & 17.9 \\
L\'erida & 127 & 24.8 & 20.3 & 29.8 & 9.9 & Gerona & 138 & 17.4 & 14.1 & 21.2 & 10.7 \\
Ja\'en & 233 & 28.8 & 25.1 & 33.0 & 7.1 & Ciudad Real & 239 & 30.5 & 26.4 & 34.4 & 6.7 \\
Las Palmas & 458 & 25.0 & 22.4 & 27.7 & 5.4 & Sevilla & 491 & 24.4 & 22.0 & 27.0 & 5.4 \\
Barcelona & 1358 & 11.1 & 10.2 & 12.1 & 4.5 & Barcelona & 1483 & 13.8 & 12.8 & 14.8 & 3.8 \\
\hline
\end{tabular*}\vspace*{30pt}
\tablewidth=\textheight
\tablewidth=\textwidth
\caption{Sample size, HB estimates of
poverty gap $\times$100, lower and upper limits of highest
posterior density intervals and coefficients of variation of HB
estimates for the Spanish provinces with sample size closest to
minimum, first quartile, median, third quartile and maximum, for
each gender}\label{TablePG}
\begin{tabular*}{\textwidth}{@{\extracolsep{\fill}}@{}ld{4.0}d{2.1}d{1.1}d{2.1}d{2.1}ld{4.0}d{2.1}d{1.1}d{2.1}d{2.1}@{}}
\hline
& \multicolumn{5}{c}{\textbf{Males}} & & \multicolumn{5}{c}{\textbf{Females}}\\[-6pt]
& \multicolumn{5}{c}{\hrulefill} & & \multicolumn{5}{c@{}}{\hrulefill}\\
\textbf{Province} & \multicolumn{1}{c}{$\bolds{n_d}$} & \multicolumn{1}{c}{$\bolds{\hat F_{1 d}^{\mathrm{HB}}}$} & \multicolumn{1}{c}{$\bolds{\mathrm{ll}(F_{1 d})}$}
                  & \multicolumn{1}{c}{$\bolds{\mathrm{ul}(F_{1 d})}$} & \multicolumn{1}{c}{$\bolds{\mathrm{cv}(\hat F_{1 d}^{\mathrm{HB}})}$} &
\textbf{Province} & \multicolumn{1}{c}{$\bolds{n_d}$} & \multicolumn{1}{c}{$\bolds{\hat F_{1 d}^{\mathrm{HB}}}$} & \multicolumn{1}{c}{$\bolds{\mathrm{ll}(F_{1 d})}$}
                  & \multicolumn{1}{c}{$\bolds{\mathrm{ul}(F_{1 d})}$} & \multicolumn{1}{c@{}}{$\bolds{\mathrm{cv}(\hat F_{1 d}^{\mathrm{HB}})}$} \\
\hline
Soria & 24 & 8.7 & 4.9 & 12.8 & 24.5 & Soria & 17 & 12.5 & 6.6 & 17.9 & 24.0 \\
L\'erida & 127 & 8.8 & 6.6 & 10.9 & 12.7 & Gerona & 138 & 5.6 & 4.3 &7.2 & 13.3 \\
Ja\'en & 233 & 10.5 & 8.4 & 12.2 & 9.3 & Ciudad Real & 239 & 10.9 & 9.1& 12.8 & 8.9 \\
Las Palmas & 458 & 8.8 & 7.6 & 10.0 & 7.0 & Sevilla & 491 & 8.2 & 7.1 &9.3 & 7.0 \\
Barcelona & 1358 & 3.3 & 2.9 & 3.7 & 5.5 & Barcelona & 1483 & 4.1 & 3.7& 4.5 & 4.8 \\
\hline
\end{tabular*}
\end{sidewaystable}

\begin{figure}[t]

\includegraphics{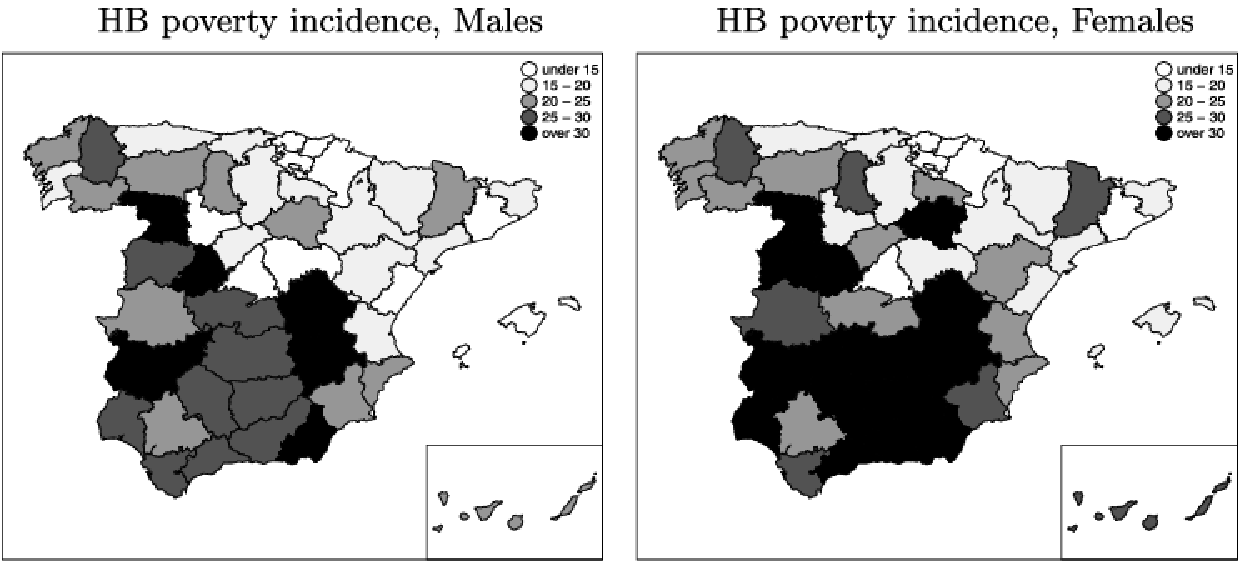}

\caption{Cartograms of estimated percent poverty
incidences in Spanish provinces for men and women,
obtained using the HB method. Canary islands have been moved to the
bottom-right corner.}\label{MapHBPI}
\end{figure}

\begin{figure}[b]

\includegraphics{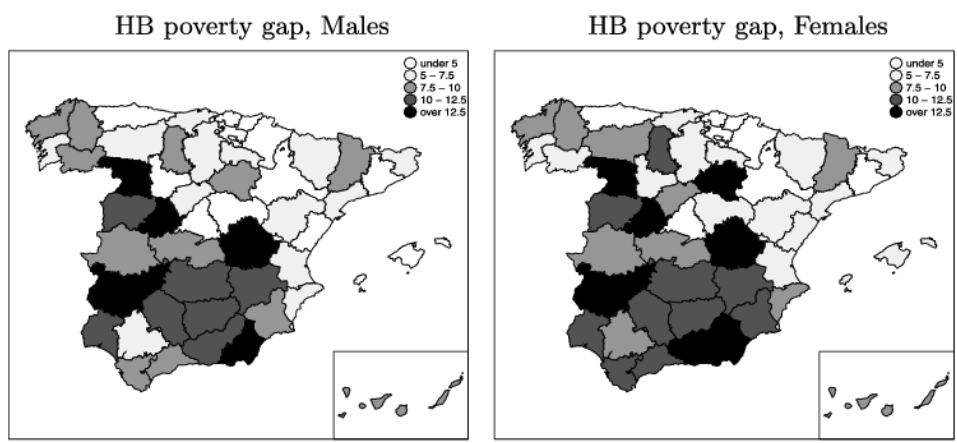}

\caption{Cartograms of estimated percent poverty gaps
in Spanish provinces for men and women,
obtained using the HB method. Canary islands have been moved to the
bottom-right corner.}\label{MapHBPG}
\end{figure}

Finally, the point estimates of poverty incidence and poverty gap
obtained using the
HB procedure are plotted in the cartograms of Figures~\ref{MapHBPI}
and \ref{MapHBPG}, for females and males.
Although the method has been applied separately for each gender in
contrast to the application done in \citet{MolRao10}
which treats provinces crossed with gender as domains, we can see that
the maps are very similar.

%s7 #&#
\section{Discussion}\label{conclusions}
The proposed HB procedure {gives efficient estimates} of general
{nonlinear} parameters
in small areas using a model for unit-level data. It is a
computationally faster alternative
to the EB method of \citet{MolRao10} and at the same time it
provides a full description
of the posterior distribution of the target parameters, making it very
easy to construct credible
intervals or to obtain other posterior summaries. The frequentist
simulation study
described in Section~\ref{FreVal} and the application with Spanish
SILC data given in Section~\ref{applic} indicate that HB
point estimates agree to a great extent with EB estimates and that
posterior variances are also
comparable with frequentist MSEs. This good property arises from the
fact of using only
noninformative priors. Thus, the proposed HB method is in practice
more feasible than the EB method for the estimation
of general nonlinear indicators under large populations.

%
%f18 #&#

%f19 #&#

{In addition, the proposed HB approach provides estimators of poverty
indicators in
Spanish provinces that are considerably more efficient than direct
estimators}; see Figure~\ref{plotCVDirecHB}. Results
highlight larger point estimates of poverty incidence and poverty gap
for females in almost all provinces
although credible intervals for the two genders cross each
other.

According to the resulting poverty maps, poverty (both in
frequency and intensity) is mainly concentrated in the south and
west of Spain. Provinces with critical estimated poverty
incidences for men, with at least 30\% of people under the poverty line,
are Almer\'{\i}a, Badajoz, Albacete, Cuenca, \'{A}vila and Zamora. For
women, almost all provinces get a larger point estimate of poverty incidence.
In particular, many more provinces join the set of provinces with
critical values, namely, practically all provinces in
the region of Andaluc\'{\i}a except for Sevilla and C\'{a}diz,
all the provinces in Castilla la Mancha region except for Toledo
and two more provinces in Castilla Le\'{o}n. Lleida in the north--east
(region of Catalonia) obtains a
worrying poverty incidence for women as compared with the rest of the provinces
in the same region. In Spain the frequency of poverty as
measured by the poverty incidence seems to be very much related
with the intensity of poverty as measured by the poverty gap,
with maps for the poverty gaps showing a similar distribution of
the poverty across provinces.

In contrast to the case of estimating area means or totals, the
proposed method,
as any other unit-level method for estimating nonlinear parameters,
requires the values of the auxiliary variables for each population
unit instead of
area aggregates. These data can be obtained from the last census or
from administrative registers.
But due to confidentiality issues and depending on the particular regulation
of each country, census data might not be easily available for
practitioners beyond
statistical offices personnel. In some other countries, access to
census data can be obtained
by prior signature of strict data protection contracts. In other small
area applications, such as, for example, agriculture or forest
research, the population of $x$-values is fully available to the researcher
from satellite or laser sensors images [\citet{BatHarFul88};
\citet{BreAst12}].

A model with spatial correlation among provinces might be
considered, but there are serious difficulties in defining
boundary conditions, especially for several provinces such as islands.
Even if spatial correlation could be
considered for a subset of the provinces, the number of provinces left
is not
large enough to estimate accurately the spatial correlation,
leading to weak significance of this parameter. An
area-level model with spatial correlation among Spanish provinces is
studied by \citet{MarMolMor13}, and their results for the SILC
data indicated very mild gains
in efficiency due to the introduction of the spatial correlation in the
model. In any case, we leave this for further research.

%sA #&#
\begin{appendix}
%sB #&#
\section{Derivation of posterior densities}\label{posterior}
Here we derive the conditional distributions appearing
on the right-hand side of the chain rule in (\ref{post}). By
Bayes' theorem and using model assumptions (\ref{condit})--(\ref{ud})
together with the prior (\ref{prior}), the posterior distribution is
given by
%
%eB.1 #&#
\begin{eqnarray}\label{postden}
&&\pi\bigl(\mathbf{u},\bolds{\beta},\sigma^2,\rho|\y_s
\bigr)\nonumber
\\
&&\qquad  \propto\Biggl\{\prod_{d=D^*+1}^D \pi
\bigl(u_d|\bolds{\beta},\sigma^2,\rho\bigr) \Biggr\}
\biggl(\frac{1-\rho}{\rho} \biggr)^{D^*/2} \bigl(\sigma^2
\bigr)^{- (((D^*+n)/2)+1)}
\\
&&\quad\qquad{}\times\exp\Biggl\{-\frac{1}{2\sigma^2}\sum
_{d=1}^{D^*} \biggl[ \sum_{i\in s_d}w_{di}
\bigl(Y_{di}-\x_{di}'\bolds{
\beta}-u_d\bigr)^2+\frac{1-\rho}{\rho}u_d^2
\biggr] \Biggr\},\nonumber
\end{eqnarray}
where $\pi(u_d|\bolds{\beta},\sigma^2,\rho)$ is the normal prior of
$u_d$ given in (\ref{ud}). Let us define the weighted sample means
\[
\bar\x_d=\frac{1}{w_{d\cdot}}\sum_{i\in s_d}w_{di}
\x_{di},\qquad\bar y_d=\frac{1}{w_{d\cdot}}\sum
_{i\in s_d}w_{di}Y_{di},
\]
where $w_{d\cdot}=\sum_{i\in s_d}w_{di}$, $d=1,\ldots,D^*$.
Integrating out $\mathbf{u}$ in (\ref{postden}), we obtain $\pi(\bolds
{\beta},\sigma^2,\rho|\y_s)$.
Now dividing $\pi(\mathbf{u},\bolds{\beta},\sigma^2,\rho|\y_s)$ by
$\pi(\bolds{\beta},\sigma^2,\rho|\y_s)$, we obtain
\[
\pi\bigl(\mathbf{u}|\bolds{\beta},\sigma^2,\rho,\y_s
\bigr)= \Biggl\{\prod_{d=D^*+1}^D \pi
\bigl(u_d|\bolds{\beta},\sigma^2,\rho\bigr) \Biggr\}
\Biggl\{\prod_{d=1}^{D^*} \pi
\bigl(u_d|\bolds{\beta},\sigma^2,\rho,\y_s
\bigr) \Biggr\},
\]
where
%
%eB.2 #&#
\begin{equation}
\label{postud} u_d|\bolds{\beta},\sigma^2,\rho,
\y_s \stackrel{\mathrm{ind}}\sim N \biggl[\lambda_d(\rho) \bigl(
\bar y_d-\bar\x_d'\bolds{\beta}\bigr),\bigl
\{1-\lambda_d(\rho)\bigr\}\frac{\rho}{1-\rho} \sigma^2
\biggr]
\end{equation}
for $\lambda_d(\rho)=w_{d\cdot}[w_{d\cdot}+(1-\rho)/\rho]^{-1}$,
$d=1,\ldots, D^*$. The second conditional density $\pi_2(\bolds{\beta
}|\sigma^2,\rho,\y_s)$ in
(\ref{post}) is obtained by integrating out $\bolds{\beta}$ in $\pi
(\bolds{\beta},\sigma^2,\rho|\y_s)$ and then dividing
$\pi(\bolds{\beta},\sigma^2,\rho|\y_s)$ by $\pi(\sigma^2,\rho|\y_s)$. Let
\begin{eqnarray*}
\mathbf{Q}(\rho)&=& \sum_{d=1}^{D^*}\sum
_{i\in s_d}w_{di}(\x_{di}-\bar
\x_d) (\x_{di}-\bar\x_d)' +
\frac{1-\rho}{\rho}\sum_{d=1}^{D^*}
\lambda_d \bar\x_d\bar\x_d',
\\
\mathbf{p}(\rho)&=& \sum_{d=1}^{D^*}\sum
_{i\in s_d}w_{di}(\x_{di}-\bar
\x_d) (Y_{di}-\bar y_d) +\frac{1-\rho}{\rho}
\sum_{d=1}^{D^*}\lambda_d \bar
\x_d\bar y_d
\end{eqnarray*}
and $\hat{\bolds{\beta}}(\rho) = \mathbf{Q}^{-1}(\rho)\mathbf{p}(\rho
)$. Then, it follows that
%
%eB.3 #&#
\begin{equation}
\label{postbeta} \bolds{\beta}|\sigma^2,\rho,\y_s\sim N
\bigl\{\hat{\bolds{\beta}}(\rho),\sigma^2\mathbf{Q}^{-1}(
\rho) \bigr\}.
\end{equation}
Finally, integrating out $\sigma^2$ in
$\pi(\sigma^2,\rho|\y_s)$, we obtain
%
%eB.4 #&#
%eB.5 #&#
\begin{eqnarray}\label{postrho}
\pi_4(\rho|\y_s)\propto\biggl(\frac{1-\rho}{\rho} \biggr)^{D^*/2}\bigl|\mathbf{Q}(\rho)\bigr|^{-1/2}\gamma(
\rho)^{-(n-p)/2}\prod_{d=1}^{D^*}
\lambda_d^{1/2}(\rho),
\nonumber\\[-12pt]\\[-4pt]
\eqntext{\varepsilon\leq\rho\leq1-\varepsilon,}
\end{eqnarray}
where
\begin{eqnarray*}
\gamma(\rho)&=&\sum_{d=1}^{D^*} \sum
_{i\in s_d}w_{di} \bigl\{Y_{di}-\bar
y_d-(\x_{di}-\bar\x_d)'\hat{
\bolds{\beta}}(\rho) \bigr\}^2
\\
&&{}+\frac{1-\rho}{\rho}\sum_{d=1}^{D^*}
\lambda_d(\rho) \bigl\{\bar y_d-\bar
\x_d'\hat{\bolds{\beta}}(\rho) \bigr\}^2.
\end{eqnarray*}
Dividing $\pi(\sigma^2,\rho|\y_s)$ by $\pi_4(\rho|\y_s)$
and making a change of variable, we finally obtain
%
%eB.6 #&#
\begin{equation}
\label{postsigmam2} \sigma^{-2}|\rho,\y_s \sim\operatorname{Gamma}
\biggl( \frac{n-p}{2}, \frac{\gamma(\rho)}{2} \biggr).
\end{equation}

%sC #&#
\section{Propriety of the posterior distribution}\label{propriety}
%
%le1 #&#
\begin{Lemma}\label{le1app2}
Under the model defined by (\ref{condit}), (\ref{ud}) and
(\ref{prior}), the posterior density $\pi(\mathbf{u}, \bolds{\beta},
\sigma^2, \rho\mid\y_s)$ is proper provided
that the matrix defined by stacking the rows~$\x_{di}'$ in columns,
$X=\mathrm{col}_{1\leq d\leq D}\mathrm{col}_{i\in s_d} (\x_{di}')$,
has full column rank and $\varepsilon\leq\rho\leq1-\varepsilon,
\varepsilon>0$.
\end{Lemma}

\begin{pf}
We need to show that $\int\!\!\int\!\!\int\!\!\int\pi(\mathbf{u},\bolds{\beta
},\sigma^2,\rho|\y_s)
\,d\mathbf{u} \,d\bolds{\beta} \,d\sigma^2\,\mathbf{d}\rho$ is finite, where
the posterior $\pi(\mathbf{u},\bolds{\beta},\sigma^2,\rho|\y_s)$
is given in (\ref{post}); see also Appendix~\ref{posterior}.

Now, using the expression for the posterior given in (\ref{post}),
the integral of the posterior distribution is given by
\begin{eqnarray*}
&& \int\!\!\int\!\!\int\!\!\int\pi\bigl(\mathbf{u},\bolds{\beta},
\sigma^2,\rho|\y_s\bigr) \,d\mathbf{u} \,d\bolds{\beta}\,\mathbf{d}\sigma^2\,\mathbf{d}\rho
\\
&&\qquad = \int\biggl[\int\biggl\{\int\biggl(\int\pi_1
\bigl(\mathbf{u}|\bolds{\beta},\sigma^2,\rho,\y_s\bigr)\,d
\mathbf{u} \biggr) \pi_2\bigl(\bolds{\beta}|\sigma^2,\rho,
\y_s\bigr)\,d\bolds{\beta} \biggr\}
\\
&&\hspace*{192pt}{}\times \pi_3\bigl(\sigma^2|\rho,\y_s\bigr)\,d\sigma^2 \biggr]
\pi_4(\rho|\y_s) \,d\rho.
\end{eqnarray*}
Here $\pi_1(\mathbf{u}|\bolds{\beta},\sigma^2,\rho,\y_s)=\prod_{d=1}^D
\pi_{1d}(u_d|\bolds{\beta},\sigma^2,\rho,\y_s)$, and the distribution of
$u_d|\bolds{\beta},\sigma^2,\rho,\y_s$ is given by (\ref{postud}),
which is
proper (integrates to one), because $\rho\in
[\varepsilon,1-\varepsilon]$ for $\varepsilon>0$. Similarly, the
distribution of $\bolds{\beta}|\sigma^2,\rho,\y_s$ is given in
(\ref{postbeta}), where the inverse of $Q(\rho)$ exists whenever
$\rho\in[\varepsilon,1-\varepsilon]$ for $\varepsilon>0$ and $X$ has full
column rank. Concerning $\sigma^2$, the density of $\sigma^{-2}|\rho,\y
_s$ is given in (\ref{postsigmam2}).
Making the change of variable $v=\sigma^{-2}$, we obtain $\int
\pi_3(\sigma^2|\rho,\y_s)\,d\sigma^2=1$.

Finally, we note that $\rho$ cannot be integrated out analytically because
the posterior of $\rho$ is given up to a constant by (\ref{postrho}). However,
\[
\int_{\varepsilon}^{1-\varepsilon} \biggl(\frac{1-\rho}{\rho}
\biggr)^{D^*/2}\bigl|\mathbf{Q}(\rho)\bigr|^{-1/2}\gamma(\rho)^{-(n-p)/2}
\prod_{d=1}^{D^*} \lambda_d^{1/2}(
\rho)\,d\rho<\infty,
\]
because the integrand is continuous for
$\rho\in[\varepsilon,1-\varepsilon]$ provided that $X$ has full column
rank.
\end{pf}

%sD #&#
\section{Computation of standardized cross-validation~residuals}\label{Appendix3}
Following \citet{GelDeyCha92}, the expectation of any function
$g(Y_{di})$ can be expressed as
%
%eD.1 #&#
\begin{eqnarray}\label{expgy}
&& E \bigl[g(Y_{di})|\y_{s(di)} \bigr]\nonumber
\\
&&\qquad = \int E \bigl[g(Y_{di})|\y_{s(di)},\btheta\bigr]\pi(\btheta|
\y_{s(di)}) \,d\btheta
\\
&&\qquad\quad = \frac{\int E [g(Y_{di})|\y_{s(di)},\btheta
]((\pi(\btheta|\y_{s(di)}))/(\pi(\btheta|\y_s))) \pi(\btheta|\y_s)
\,d\btheta}{\int
((\pi(\btheta|\y_{s(di)}))/(\pi(\btheta|\y_s))) \pi(\btheta|\y
_s)\,d\btheta}.\nonumber
\end{eqnarray}

Now, to obtain the expectation $E (Y_{di}|\y_{s(di)} )$,
consider $g(x)=x$. The expectation within the integral in (\ref{expgy}) is
simply
\[
E (Y_{di}|\y_{s(di)},\btheta)=E (Y_{di}|\btheta)=\x
_{di}'\bolds{\beta}+u_d,
\]
because, given $\btheta$, all observations are independent and
distributed as indicated in (\ref{predfdi}). Thus, if we generate $H$ values
$\btheta^{(h)}= ((\mathbf{u}^{(h)})',(\bolds{\beta}^{(h)})',\sigma
^{2(h)},\rho^{(h)})'$, $h=1,\ldots,H$, from the posterior
density with all the data, $\pi(\btheta|\y_s)$, then the desired
expectation would be
obtained as
\[
E (Y_{di}|\y_{s(di)} )\approx\sum_{h=1}^H
\bigl(\x_{di}'\bolds{\beta}^{(h)}+u_d^{(h)}
\bigr)v_{di}^{(h)},\qquad i\in s_d, d=1,\ldots,D,
\]
where
\[
v_{di}^{(h)}= \Biggl\{\sum_{k=1}^H
\frac{\pi(\btheta^{(k)}|\y
_{s(di)})}{\pi(\btheta^{(k)}|\y_s)} \Biggr\}^{-1} \frac{\pi(\btheta
^{(h)}|\y_{s(di)})}{\pi(\btheta^{(h)}|\y_s)}.\vadjust{\goodbreak}
\]
But by Bayes' theorem, we have
\[
\frac{\pi(\btheta^{(h)}|\y_{s(di)})}{\pi(\btheta^{(h)}|\y_s)}=\frac{f(\y
_{s(di)}|\btheta^{(h)})}{f(\y_s|\btheta^{(h)})}\frac{f(\y_s)}{f(\y_{s(di)})}.
\]
Therefore,
\begin{eqnarray*}
v_{di}^{(h)}&=&\frac{((f(\y_{s(di)}|\btheta^{(h)}))/(f(\y
_s|\btheta^{(h)})))((f(\y_s))/(f(\y_{s(di)})))}{\sum
_{k=1}^H ((f(\y_{s(di)}|\btheta^{(k)}))/(f(\y_s|\btheta^{(k)})))
((f(\y_s))/(f(\y_{s(di)})))}
\\
&=& \frac{f(\y_{s(di)}|\btheta^{(h)}) \{f(\y
_{s}|\btheta^{(h)})
\}^{-1}}{\sum_{k=1}^H f(\y_{s(di)}|\btheta^{(k)}) \{
f(\y_{s}|\btheta^{(k)}) \}^{-1}}.
\end{eqnarray*}
Now since, given $\btheta$, all observations are independent, we
have $f(\y_s|\btheta)=f(\y_{s(di)}|\btheta)f(Y_{di}|\btheta)$.
Replacing this relation in $v_{di}^{(h)}$, we obtain the expression in~(\ref{vdi}).
\end{appendix}

\section*{Acknowledgments}
We thank the referees and the Associate Editor for constructive
comments and suggestions.

%suskaldyti doi

% imsref loaded by linak, 2014-01-22 09:12:28
% imsref loaded by linak, 2014-01-22 09:18:01
% imsref loaded by linak, 2014-01-22 09:19:57
% imsref loaded by linak, 2014-01-22 09:23:04
% imsref loaded by linak, 2014-01-22 09:35:30
% imsref loaded by linak, 2014-01-22 09:45:50
%
% imsref loaded by linak, 2014-05-30 12:09:40

\printaddresses

\end{document}